\documentclass[prd,aps,twocolumn,floatfix]{revtex4}

\usepackage{amssymb,graphicx}

\newcommand{\be}{\begin{equation}} 
\newcommand{\ee}{\end{equation}} 
\newcommand{\bea}{\begin{eqnarray}} 
\newcommand{\eea}{\end{eqnarray}} 
\newcommand{\beay}{\begin{eqnarray*}}

\newcommand{\eps}{\varepsilon}

\begin{document} 
\renewcommand{\arraystretch}{2.} 
\setlength{\tabcolsep}{3mm} 
 
%\input psfig 
%\pssilent 
\title{The discrete energy method in numerical relativity: Towards long-term 
stability} 

\author{Luis Lehner$^1$, David Neilsen$^1$, Oscar Reula$^2$, and Manuel
  Tiglio$^1$}
\altaffiliation{Current address: 78 Bull Hill Road, Newfield, New York 14867.}
\affiliation{$1$ Department of Physics and Astronomy, Louisiana State
University, Baton Rouge, LA 70803-4001\\
$2$ FaMAF, Universidad Nacional de C\'ordoba, C\'ordoba, Argentina 5000}

\begin{abstract} 
The energy method can be used to identify well-posed initial boundary 
value problems for quasi-linear, symmetric hyperbolic partial differential 
equations with maximally dissipative boundary conditions.  A similar 
analysis of the discrete system can be used to construct stable 
finite difference equations for these problems at the linear level.  In this paper we 
apply these techniques to some test problems commonly used in numerical 
relativity and observe that while we obtain convergent schemes, 
fast growing modes, or ``artificial instabilities,'' contaminate 
the solution.  We find that these growing modes  
 can partially arise from 
the lack of a Leibnitz rule for discrete derivatives and discuss ways 
to limit this spurious growth. 
\end{abstract} 
 
\maketitle 
 
\section{Introduction} 
 
Einstein's theory of general relativity is described by a  
complicated set of coupled, nonlinear partial differential  
equations.  Like the Maxwell equations of 
classical electromagnetism, the Einstein equations are over-determined, 
and can be separated into hyperbolic evolution and elliptic 
constraint equations, as in the Arnowitt--Deser--Misner (ADM)  
decomposition~\cite{ADM}.  The complexity of these equations 
is such that analytic solutions have been found for only very special 
configurations, and numerical studies arguably provide the only means to 
explore a wide variety of astrophysical and theoretically significant 
problems~\cite{lehnerreview,shapirobaumgarte}.  Unfortunately, these 
numerical solutions are often prone to various instabilities, which we 
divide into three classes: (1) continuum instabilities, (2) numerical 
instabilities and (3) artificial long-term ``instabilities'' (ALTI). 
 
{\it Continuum instabilities} exist in the formulation of the continuum 
problem (and thus are naturally reflected in any consistent 
numerical scheme), and are characterized by the solution itself or some
perturbation of it  
either ``blowing up'' at a finite time or growing fast in time. These instabilities may reflect physical 
phenomena, such as turbulence in fluids or the threshold of black hole 
formation~\cite{gundlach}, or may be an artifact of the 
formulation of the problem. For instance, ill-posed problems suffer from
these instabilities, 
which also manifest themselves as numerical ones  
(see ~\cite{gko}, and~\cite{convergence} for an illustration in 
numerical relativity). {\it Numerical instabilities}  are characterized  
by errors that, at a fixed time, {\em increase} as the discretization scale is decreased.   
These instabilities are present in many discretization schemes for well-posed problems that appear 
``natural'', such as the forward-time, centered-space scheme for the 
advection equation.   Finally, {\it ALTI} are 
sometimes exhibited even by numerically {\it stable} implementations, whenever errors grow 
 too fast for the desired simulation time-scale and available computational 
resources.  Although these may  
not be instabilities in the strict sense, given that --as defined here-- they are not 
present at the continuum and do go away with resolution, they  
are problematic in practice 
as they limit the time-length of a reliable simulation.  In this paper we 
focus on this long-term error growth together with the application of 
the techniques described in~\cite{paper1}, which throughout this 
work we refer to as Paper~I. 
 
The stability of finite difference approximations (FDAs) for simple 
Initial Value Problems (IVP) with constant coefficients and periodic 
boundaries is often analyzed in discrete Fourier space, by examining the 
norm of the amplification matrix.  This method, however, does not generalize
in a straightforward way 
to more complicated problems, such as those with variable 
coefficients or those with boundaries.  A more powerful technique is 
based on the energy method~\cite{gko}, also used to identify well-posed 
continuum problems. This method can be used in all problems for which the continuum energy 
method applies, e.g., quasi-linear parabolic and symmetric hyperbolic 
partial differential equations with appropriate boundary conditions. 
The discrete formulation reproduces the continuum analysis, 
including the calculation of a discrete energy estimate, which 
can be used to meet sufficient conditions for a stable numerical method. 
Combining these estimates with the consistency of the discrete system with 
the continuum problem, stability of the linear problem is equivalent 
to convergence via the Lax equivalence theorem~\cite{Lax}.   
While our focus in general relativity is on the nonlinear Einstein equations,  
 stability of the linearized equations is obviously a necessary step  
for solving non-linear problems, in particular those with smooth solutions.  
In the absence of matter, 
smooth solutions are indeed expected in general relativity when appropriate 
slicing conditions are used~\cite{hyp}.
 
As described in Paper~I, a discrete energy estimate can be obtained 
when using finite difference operators that satisfy {\it summation 
by parts} (SBP)~\cite{kreiss-scherer}, a discrete version of integration 
by parts, as well as appropriate boundary conditions~\cite{olsson} 
and time integrators~\cite{kreiss-wu,tadmor}.  Paper~I presents and extends  
these techniques in the context of numerical relativity.  In this paper we 
demonstrate these techniques using some common test problems employed in the field. 
 We evolve Klein--Gordon scalar fields and Maxwell fields 
on fixed Schwarzschild spacetimes. We also consider gauge wave spacetimes,
obtained via coordinate transformations of flat spacetime. We show that we naturally obtain 
stable and convergent numerical schemes by construction, even when an 
inner boundary is used to remove the black hole singularity (black hole 
excision)~\cite{unruh}.
 
A second goal of this paper is to understand some causes of long-term 
error growth (ALTI), and ways of eliminating or minimizing it by achieving 
{\it strict stability}~\cite{olsson}.  Numerical solutions of the Einstein 
equations are frequently plagued by such long-term errors. Since at least some 
of these appear in numerically stable implementations, their cause is often attributed 
to the formulation of the equations. While this may be the case in a 
number of problems, as discussed below, some numerical experiments of 
simple differential equations with no continuum instabilities also display this same artificial 
long-term instability ---even when discretized in a numerically stable way---.  Hence,  it  
is important to devise {\em numerical} techniques to 
control this spurious growth in the numerical error.  We find that this 
spurious error growth is allowed in the absence of sharp energy estimates 
for the discrete problem.  
This can arise, from among other reasons, 
because the discrete derivatives fail to satisfy the Leibnitz Rule. 
With this observation in mind, we discuss here a method to remedy some of these problems by constructing 
schemes that suppress excitations of ALTI. 
 
This work is organized as follows:  We first summarize, in Section II, some results from 
Paper~I, in order to introduce some notation.  Section~\ref{strict} 
discusses the notion of strict stability, and we present simple examples 
of ``natural'' schemes that are numerically stable but not strictly 
stable, for which the errors grow fast in time.  We then explain the 
basic strategy to achieve strict stability, and show 
how this eliminates ALTI. In Section~\ref{equations} we present some test problems from 
numerical relativity: three-dimensional scalar and electromagnetic 
fields propagating on a fixed Schwarzschild black hole geometry,  and 
one-dimensional perturbations of a dynamical slicing of flat spacetime 
(``gauge wave'').  The first cases allow us to test black hole excision, while the  
gauge wave has a time dependent background without a sharp energy 
estimate.  However, a sharp energy estimate does exist for the subsidiary 
constraint system, and it can be 
used to understand the nature of the ALTI and suppress them. Although 
more work would need to be done, these observations and ideas could be 
useful when evolving Einstein's equations in more complicated situations. 
In Section~\ref{simulations} we present numerical simulations of the 
systems discussed in Section~\ref{equations}. 
Finally, in appendix~\ref{smallness} we discuss   
cubical excision boundaries for black hole spacetimes. 
 
%%%%%%%%%%%%%%%%%%%%%%%%%%%%%%%%%%%%%%%%%%%%%%%%%%%%% 
\section{Numerical stability and the energy method: an overview}  
\label{sbp} 
%%%%%%%%%%%%%%%%%%%%%%%%%%%%%%%%%%%%%%%%%%%%%%%%%%% 
 
We first review the construction of numerically stable finite-difference schemes using the  
discrete energy method for linear, symmetric hyperbolic  
initial-boundary value problems (IBVPs) with maximally dissipative boundary 
conditions.  While the discussion here 
is brief, further information can be found in the standard text by 
Gustaffsson, Kreiss, and Oliger~\cite{gko}, works by Olsson~\cite{olsson}, 
and Paper~I, for their particular application in numerical relativity. 
Sufficient conditions for a stable discretization are fulfilled when: 
\begin{enumerate} 
\item The continuum problem can be shown to be well posed using the energy method. 
\item Spatial difference operators  
      that satisfy SBP on the computational domain are
      used. 
\item Boundary conditions are imposed via orthogonal projections~\cite{olsson}. 
\item The semi-discrete equations are integrated with a {\em locally 
      stable} time integrator, e.g., third or fourth order  
      Runge--Kutta~\cite{kreiss-wu,tadmor}. 
\item Optionally, explicit dissipation may be added using operators 
      that do not spoil the energy estimates, and different ways of writing
      the discrete equations may be explored to achieve strict stability.  
\end{enumerate} 
 
As an example, consider a set of linear, symmetric hyperbolic equations  
on a one-dimensional domain, $x \in [a,b]$.  This domain is discretized  
with points $x_i= a + i\Delta x$, $i=0\ldots N$, where $\Delta x=(b-a)/N$, 
and a discrete scalar product,  
\begin{equation} 
(u,v) := \Delta x \sum_{i,j=0}^N \sigma_{ij} u_i v_j \, , 
\end{equation} 
is introduced for some positive definite matrix with elements $\sigma_{ij}$. 
In the limit  $\Delta x \to 0$ this scalar 
product approaches the continuum one given by 
\begin{equation} 
\langle u,v\rangle :=  \int_a^b uv \,dx \,. 
\end{equation} 
The standard continuum derivative operators and scalar products satisfy 
integration by parts: 
\begin{equation} 
\langle u',v\rangle  + \langle u,v'\rangle  = uv|_a^b \label{ibp}. 
\end{equation} 
Similarly, a discrete difference operator $D$ is said to satisfy SBP, 
if there is a scalar product with respect to which 
\begin{equation} 
(Du,v)  + (u,Dv)  = uv|_a^b
\end{equation} 
holds. The simplest difference operator and scalar product that satisfy SBP on this 
domain are  
\begin{equation} 
\begin{array}{lll} 
D = D_+\,,\quad & \sigma_{00}=\frac{1}{2}\quad & \mbox{ for } i=0\\ 
D = D_0\,,\quad & \sigma_{ii}=1          \quad & \mbox{ for } i=1\ldots N-1\\ 
D = D_-\,,\quad & \sigma_{NN}=\frac{1}{2}\quad & \mbox{ for } i=N 
\end{array} \label{simpled}
\end{equation} 
where $D_0u = (u_{i+1}- u_{i-1})/(2\Delta x)$,  
$D_- u = (u_{i}-u_{i-1})/\Delta x$, $D_+ u = (u_{i+1}-u_{i})/\Delta x$, and
the scalar product is diagonal: $\sigma_{ij}=0$ for $i\ne j$. 
Higher order operators satisfying SBP have been constructed by 
Strand~\cite{strand}.  Domains with inner boundaries in several dimensions 
introduce additional complexities, and we refer the reader to Paper~I for 
more information. 
 
We sometimes use explicit numerical dissipation. For that purpose we modify near
boundaries the standard Kreiss--Oliger dissipative operator for second order derivatives, $Q_d$, so that it does 
not spoil the semi-discrete estimate, i.e., so that it satisfies $(u,Q_du)\le 0$.  
The operator $Q_d$, which is fully described in Paper~I, is added to the
right-hand side of the evolution equations. In the absence of inner
boundaries, it is (on each direction)   
\begin{eqnarray}  
Q_d u_0 &=& -2\epsilon \Delta x D_+^2 u_0,\nonumber\\  
Q_d u_1 &=& -\epsilon \Delta x (D_+^2 -2D_+D_-)u_1,\nonumber\\  
Q_d u_i &=& -\epsilon (\Delta x)^3(D_+D_-)^2 u_i, 
           \; \mbox{ for } i=2,\ldots,N-2 \nonumber\\  
Q_d u_{N-1} &=& -\epsilon \Delta x (D_-^2 -2D_+D_-)u_{N-1}, \nonumber\\  
Q_d u_N &=& -2\epsilon \Delta x D_-^2 u_N. \label{eq:KOdiss21}  
\end{eqnarray}  

We discretize the equations in time using the method of lines, and use 
third-order Runge--Kutta to integrate the equations.  More precisely, 
for the ordinary differential system 
$$ 
  \frac{\partial u}{\partial t} = L(t,\, u), 
$$ 
we integrate the solution $u^n$, defined at time $t^n$, to the advanced 
time, $t^{n+1}=t^n+\Delta t$, obtaining $u^{n+1}$, through 
\begin{eqnarray} 
u^{(1)} &=& u^n + \frac{1}{2}\Delta t\, k_1,\nonumber\\ 
u^{(2)} &=& u^n + \frac{3}{4}\Delta t\, k_2,\nonumber\\ 
u^{n+1} &=& u^n + \frac{1}{9}\Delta t\, (2k_1 + 3k_2 + 4k_3), 
\label{eq:rk3} 
\end{eqnarray} 
where 
\begin{eqnarray} 
k_1 &=& L\left(t^n,\, u^n\right),\nonumber\\ 
k_2 &=& L\left(t^n + \frac{1}{2}\Delta t, \, u^{(1)}\right),\nonumber\\ 
k_3 &=& L\left(t^n + \frac{3}{4}\Delta t, \, u^{(2)}\right). 
\end{eqnarray} 
 
%%%%%%%%%%%%%%%%%%%%%%%%%%%%%%%%%%%%% 
\section{Artificial long-term instabilities and strict stability}  
\label{strict} 
%%%%%%%%%%%%%%%%%%%%%%%%%%%%%%%%%%%%%% 
 
Throughout this paper we call artificial long-term instabilities (ALTI) those instabilities that are  
{\it neither} numerical---the errors at a fixed time do not become larger 
when the resolution is increased, but rather converge to zero---% 
{\it nor} true instabilities of the continuous solution. These  
instabilities, as shown in some examples below, can appear quite  
generically in long-term simulations. 
 
There are least two possible ways to suppress ALTI's: 
\begin{enumerate} 
\item Constructing {\it strictly stable} discretizations with respect to  
a conveniently defined energy.  The equations are written such that a 
sharp continuum estimate also holds for the semi-discrete equations.  
The semi-discrete energy will then closely mirror the continuum energy  
estimate. 
 
\item Explicit numerical dissipation can be used to eliminate high 
frequency errors, shifting the spectrum of the amplification matrix.   
\end{enumerate} 
 
Although the latter solution is easy to implement, it may not be 
effective if the ALTI excites low frequency modes.  The first solution, 
on the other hand, does not introduce artificial damping, but such a 
discretization may be difficult or even impossible to find for generic 
systems of equations.  For example, no general sharp estimate for the  full 
Einstein equations is known at this time.  Thus, these two 
possibilities can be seen as complementary to each other. 
 
The energy of a solution, $u$,  of a set of linear \footnote {In order to
  obtain an estimate in the
  non-linear case, one might need to include derivatives of the fields in the
  energy as well.} partial differential equations is  
defined as 
\begin{equation} 
{\cal E}(t) = \langle u,  H u \rangle\;, 
\end{equation} 
where $H$ is a positive definite symmetrizer for the system. 
This energy \footnote{Despite the name, this energy {\it does not} need to correspond to the 
physical energy of the system.} is not unique, reflecting the freedom in  
choosing the symmetrizer, $H$.  
The energy method provides an estimate for the growth rate of the 
energy in the form of an inequality, say,  
\begin{equation} 
\frac{d{\mathcal E}}{dt}  \leq F( {\mathcal E}) \, . 
\end{equation} 
and thus provides a bound to the energy growth.  This bound may be much larger than the 
actual growth rate for the continuum solution, in which case we say that 
the energy estimate is {\it coarse.} When the estimate is close  
to the actual energy growth rate of a solution, the estimate is {\it 
sharp.}  While the behavior of the continuum solution is independent of a 
particular bound, as we discuss in this paper, numerical solutions sometimes do have a
large growth if this is allowed.  Thus, the sharper the bound is, the smaller the 
possible spurious growth in the numerical solution.
 
Finding a sharp estimate is  problem-dependent, and the 
definition of a convenient energy has to be analyzed on each case. 
One usually attempts to find a sharp energy bound by exploiting the freedom 
in choosing the symmetrizer.  For example, a well-defined local physical energy 
exists in many physically motivated problems, which can be used to 
select a preferred symmetrizer.  The evolution equations must be 
written such that they are symmetrizable with respect to this energy, 
but this is in general possible.  Although a locally positive definite energy for the 
Einstein equations does not exist, relatively sharp estimates exist for some 
problems~\cite{friedichstablemink,chrisklain,rodklain}. 
 
The energy method can be used to  derive discrete estimates.  Using difference operators 
that satisfy SBP and representing the boundary conditions through orthogonal
projections, 
the semi-discrete calculation mirrors the continuum 
analysis.  When a discrete energy estimate is present, 
the discrete system is numerically stable by construction. 
A discrete system with the same energy estimate as 
the continuum problem may be defined as strictly stable.  However, we are  
particularly interested in systems for which the energy estimate 
is sharp, and we therefore reserve the term {\it strictly stable} for 
discrete systems that match a sharp continuum energy estimate. 
 
Assuming a sharp energy estimate exists for the continuum problem,  
in this paper we consider how the discrete problem might be implemented to preserve 
this sharp estimate at the numerical level.  This can be exploited to achieve  
long-term stability in an 
otherwise numerically stable but long-term unstable scheme. Throughout 
this paper we only use numerically stable schemes, even when not explicitly 
mentioned.  Therefore, we concentrate on their long-term stability.  
 
Finally, we note that the sharp energy estimates discussed here are for 
semi-discrete systems, where time is continuous but space has been 
discretized.  A sharp estimate may, in principle, be lost owing to  
error introduced by the temporal integration \footnote{Though some energy estimate is
still available, as discussed in Paper I, and the scheme is at least
numerically stable.}.  The experiments presented  
in this paper and in \cite{hyperGR} suggest that (at least in the cases considered) if 
strict stability  is lost, the effects are not serious.  The semi-discrete 
analysis appears to adequately represent the fully discrete simulation.

%%%%%%%%%%%%%%%%%%%%%%%%%%%%%%%%%% 
\subsection{Example 1: A simple model} 
%%%%%%%%%%%%%%%%%%%%%%%%%%%%%%%%% 
 
\begin{figure}[ht] 
\begin{center} 
\includegraphics*[height=6cm]{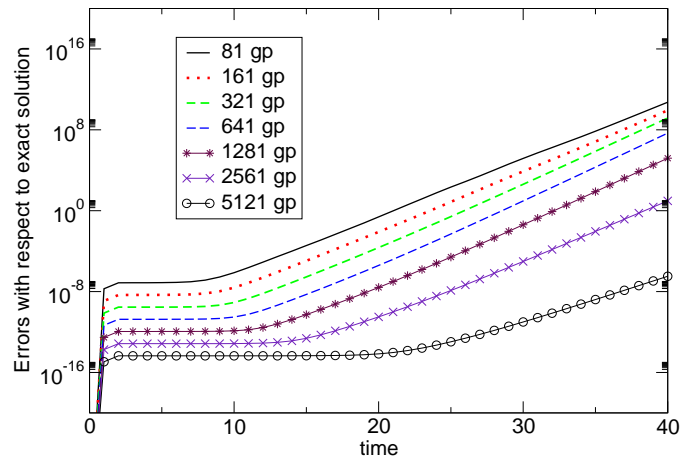} 
\caption{This figure shows the $L_2$ norm of the solution error for  
Eq.~(\ref{1da}) as a function of time at many different discretization 
scales, with initial data $u(0,x)=1/x$ and boundary conditions $u(t,2)=1/2$.  
The solution is calculated with a numerically stable 
discretization that is not strictly stable.  Although convergent, the 
error in the solution clearly displays an exponential growth. The CFL factor used is 0.8. 
}. 
\label{conv_simple_eq}  
\end{center} 
\end{figure}

Consider the initial-boundary value problem defined by the equation 
\begin{equation} 
\dot{u}= u' + u/x \; , \qquad x\in [1,2], \qquad t\ge 0 \;, \label{1d} 
\end{equation} 
together with some initial data and maximally dissipative boundary conditions.  
Let us consider the numerically stable discretization defined by 
\begin{equation} 
\dot{u}_i = D u_i + u_i/x_i  \;\;\;, \;\;\; i=0\ldots N, \qquad t\ge 0 \;, 
\label{1da}  
\end{equation} 
with the discrete derivative, $D$, given by Eq.~(\ref{simpled}), third-order
Runge--Kutta for the time integration [cf. Eq.~(\ref{eq:rk3})], and maximally
dissipative boundary conditions imposed through orthogonal projections.  
Figure \ref{conv_simple_eq} shows results from simulations 
using this numerical scheme and $u(t=0,x)=1/x$ as initial data and
$u(t,2)=1/2$ 
as boundary conditions, which corresponds to the time-independent
solution $u(t,x)=1/x$ .  The figure plots the $L_2$ norm of 
the errors with respect to the exact solution as a function of time, 
at different resolutions (obtained on uniform grids, where the number 
of points ranges from  $81$ to $5121$). At a fixed time, 
the errors decrease with resolution, verifying that, as expected, 
the scheme is numerically stable. At fixed resolution, however, the  
error grows as a function of time, i.e., the solution has long-term 
error growth, or ALTI.

To understand this behavior we examine an energy norm of the solution. 
Let the continuum energy \footnote{The superscript index ${(1)}$ is used to denote that this is  
the first energy defined for this system.} be (using $H=1$)
\begin{equation} 
{\cal E}^{(1)} = \langle u,u\rangle  \label{cenergy1} \;,  
\end{equation} 
We obtain an energy estimate by taking a time derivative in both sides 
of Eq.~(\ref{cenergy1}) and using the evolution equation (\ref{1d}), 
obtaining 
\begin{eqnarray} 
\dot{{\cal E}}^{(1)}  & =&   \langle \dot{u},u\rangle  + \langle
u,\dot{u}\rangle \nonumber \\
&=&  \langle u'+u/x,u\rangle  + \langle u,u'+u/x\rangle  \nonumber \\
& =& \langle u',u\rangle  + \langle u,u'\rangle  +  
\langle u/x,u\rangle  + \langle u,u/x\rangle  \label{before} 
\\ 
&=& -u(a)^2 + u(b)^2 +  \langle u/x,u\rangle  + \langle u,u/x\rangle   
        \label{after} \\ 
&=&  -u(a)^2 + \frac{1}{4} +  2  \langle u/x,u\rangle \nonumber \\ 
& \le &   \frac{1}{4} + 2\langle u/x,u\rangle  \nonumber \\ 
&\le & \frac{1}{4} +  2a^{-1}\langle u,u\rangle \nonumber
\end{eqnarray} 
That is, 
\begin{equation}
\dot{{\cal E}}^{(1)} \le \frac{1}{4} +  2a^{-1}{\cal E}^{(1)} \; .\label{bad_estimate} 
\end{equation}
Only integration by parts was used to derive this estimate,  between 
Eqs.~(\ref{before}) and (\ref{after}). 
 
Now consider the discrete problem. We define a semi-discrete 
energy analogous to that one defined in Eq.~(\ref{cenergy1}), namely 
\begin{equation} 
%%%E^{(1)} = \langle u,u\rangle  \label{denergy1} \;. 
E^{(1)} = (u,u)  \label{denergy1} \;. 
\end{equation} 
The derivation of the above continuum  energy estimate 
can be repeated at the semidiscrete level, 
provided the difference operator satisfies SBP, since only
integration by parts was used at the continuum. 
Therefore, the same estimate holds for the semi-discrete energy defined  
in Eq.~(\ref{denergy1}). That is,   
\begin{equation} 
\dot{E}^{(1)} \le  \frac{1}{4} + 2a^{-1}E^{(1)}  \; . 
\label{bad_destimate} 
\end{equation} 
 
The continuum estimate, Eq.~(\ref{bad_estimate}), and the semi-discrete  
counterpart, 
Eq.~(\ref{bad_destimate}), guarantee that the norm of the continuum and numerical solutions 
will grow with a rate bounded by $e^{2t/a}$.  As these are estimates, this does not mean 
that the exponential growth actually occurs. Indeed, 
the continuum solution does not grow, since Eq.~(\ref{1d})  
is the advection equation with the time-independent change of variables, 
$v=xu$. With this transformation, Eq.~(\ref{1d}) becomes $\dot{v} = v'$. 
However, the results shown in Fig.~\ref{conv_simple_eq}  
indicate that an exponential growth does appear in the numerical solution. 
 
We now define a new energy \footnote{This time denoted with the superscript ${(2)}$.}, that allows
for a  sharper energy estimate 
and, consequently, a better discrete system.  Let this new energy  
(defined with $H=x^2$) be  
\begin{equation} 
{\cal E}^{(2)}  = \frac{1}{2}\langle u,x^2u\rangle  \;.  \label{cenergy2} 
\end{equation} 
As before, an estimate is obtained by differentiating in time, substituting 
in the evolution equations, and using integration by parts to get 
\begin{eqnarray} 
\dot{{\cal E}}^{(2)}  &=& \int_a^b x^2u\dot{u} \, dx 
= \int_a^b x^2u\left(u' + \frac{u}{x}\right)  dx \nonumber \\  
& = & \int_a^b \left( \frac{x^2u^2}{2}\right)' dx 
= -\frac{a^2u(a)^2}{2} \le 0 \; . \label{good_estimate} 
\end{eqnarray} 
In deriving the estimate (\ref{good_estimate}) we have used the Leibnitz rule, in the third equality. 
The analogous semi-discrete energy is now 
\begin{equation} 
E ^{(2)}  = \frac{1}{2}(u,x^2 u)\;.  \label{denergy2} 
\end{equation} 
However this semi-discrete energy will not, in general, satisfy a discrete
version of the continuum estimate given in (\ref{good_estimate}). The reason is that 
although $D$ satisfies SBP, it does {\em not} generally 
satisfy the Leibnitz rule,  
which was necessary in deriving the estimate (\ref{good_estimate}). It can actually be checked
that,  
because of this, discretizing the system as in Eq.~(\ref{1da}) will {\em not}  
reproduce the continuum estimate, i.e. 
\begin{equation} 
\dot{E}^{(2)} \neq -\frac{a^2u(a)^2}{2}. 
\label{good_destimate} 
\end{equation} 
An alternative is to rewrite the evolution equation (\ref{1d}) 
such that the Leibnitz rule is unnecessary to obtain the continuum 
estimate, Eq.~(\ref{good_estimate}). Then a semi-discrete 
version of this estimate does hold if the difference operator satisfies SBP.
To do so we write Eq.~(\ref{1d}) as  
\begin{equation} 
\dot{u} = \frac{1}{x}(ux)'\;. 
\label{rearrange1} 
\end{equation} 
This gives the same estimate as before: 
\begin{equation} 
\dot{{\cal E}}^{(2)}  = \int_a^b x^2u\dot{u}\,dx = \int_a^b xu (xu)'\,dx = 
- \frac{1}{2}\left[ au(a) \right]^2 \le 0. \label{est1} 
\end{equation} 
However, in deriving the estimate (\ref{est1}) we have only used integration by parts,  
and not the Leibnitz rule. 
Therefore, the semi-discrete energy estimate
\begin{equation} 
\dot{E}^{(2)} = -\frac{a^2u(a)^2}{2}. 
\end{equation}
follows immediately if the semidiscrete equation is written as
\begin{equation} 
\dot{u} = \frac{1}{x}D(ux). \label{1dc} 
\end{equation} 
and $D$ satisfies SBP. 
 
\begin{figure}[ht] 
\begin{center} 
\includegraphics*[height=6cm]{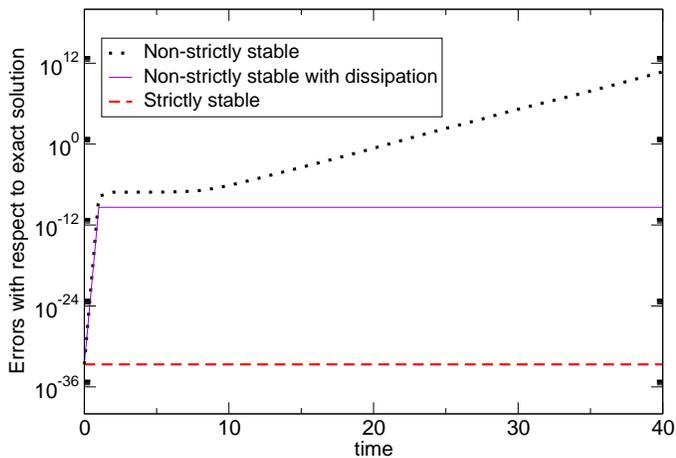} 
\caption{ 
This figure compares the solution error, at fixed resolution, for different discretizations of 
Eq.~(\ref{1d}).  The dotted line shows the one of the lines of Fig.1, i.e.,
the $L_2$ norm of the error 
for a non-strictly stable discretization, given by Eq.~(\ref{1da}). 
The solid line shows the error for the same simulation with some numerical
dissipation ($\sigma=0.01$).  
The dashed line, in turn, shows the error for the (non-dissipative) strictly
stable discretization given by  
Eq.~(\ref{1dc}).  In all these runs $\Delta x = 1/80$ and CFL=0.8. Even though
both the strictly stable and the non-strictly but dissipative schemes give
long term stability, the former discretization is much more accurate.}  
\label{comparison_discretizations}  
\end{center} 
\end{figure}

Figure~\ref{comparison_discretizations} shows numerical results obtained 
by evolving Eq.~(\ref{1d}) using two discretizations, given by   
Eqs.~(\ref{1da},\ref{1dc}). 
Eqs.~(\ref{1d}) and (\ref{rearrange1})  are equivalent, they have the same set of solutions. 
However, their discrete counterparts,  
Eqs.~(\ref{1da}) and (\ref{1dc}), {\em do not} have the same solutions  
at fixed resolution.  
In one case the solution error 
grows as a function of time, in the other case it does 
not.  

It is experimentally found that in this example (and others discussed below) 
the growing numerical error in the non-strictly stable discretization 
is manifest in high frequency modes,  
which can be eliminated by adding a small amount of 
dissipation, as is also shown in Figure~\ref{comparison_discretizations}.

%%%%%%%%%%%%%%%%%%%%%%%%%%%%%%%%%%%%%%%%% 
\subsection{Example 2: Energy preserving and non-preserving system} 
%%%%%%%%%%%%%%%%%%%%%%%%%%%%%%%%%%%%%%%% 
 
Consider now an equation with variable-coefficients in the principal part and 
a lower order term that in principle allows for energy growth,  
\begin{equation} 
\dot{u} = 2cu' + u c' + F(u),  \label{shift_eq}
\end{equation} 
defined on a compact domain $x \in [a,b]$. For simplicity we assume that 
$c=c(x)>0$, and define the energy ${\cal E}=\langle u,u \rangle/2$.  
Homogeneous boundary conditions are given at $x=b$. 
Rewriting the equation as $\dot{u} = (uc)' + cu' + F(u)$, one 
can easily show, using only integration by parts, that the estimate 
$$ 
\dot{{\cal E}} = (u^2c)|^b_a + \langle F,u \rangle = -u(a)^2c(a) + \langle F,u \rangle 
\le \langle F,u \rangle   
$$ 
holds. Therefore, writing the semi-discrete equation as  
\begin{equation}
\dot{u} = D(uc) + cDu + F(u) \label{discr}
\end{equation}
gives the same semi-discrete energy estimate, 
$$ 
\dot{E}  \le ( F,u )
$$ 
If $F=0$ then $\dot{E},\dot{{\cal E}} \le 0$ and the continuum and discrete solutions cannot grow 
in time.  Equations with features similar to those of
Eq.(\ref{shift_eq}) with $F=0$ 
appear in the  Klein--Gordon and Maxwell fields on fixed black hole backgrounds, as  
discussed later in this paper,  where the shift plays the role of $c$. 
 
If $F\ne 0$ the energy may grow, already at the continuum; for example, if $F=u$. 
In that case one may  want to discretize in a way such that artificial numerical growth 
is avoided as much as possible.  For example, if $F=u$, discretizing as in Eq.~(\ref{discr}) gives
$$
\dot{E} =   -u^2(a)c(a) +  E \;,
$$
which is the same equation satisfied by the continuum energy.

%%%%%%%%%%%%%%%%%%%%%%%%%%%%%%%%%%%%%%%%% 
\subsection*{Summary} 
%%%%%%%%%%%%%%%%%%%%%%%%%%%%%%%%%%%%%%%% 
 
Fast growing numerical errors can arise even in numerically stable 
discretizations. In some cases they can be traced to the failure of the 
Leibnitz rule for discrete derivatives, and failure of sharp continuum 
energy estimates to hold in the discrete problem. The use of difference 
operators satisfying SBP, 
appropriate numerical boundary conditions, and rearrangement of 
terms, give raise to strictly stable schemes 
which satisfy sharp semidiscrete energy estimates, if such estimates are 
available at the continuum.

The goal of these rearrangement of terms is actually not only to avoid the Leibnitz
rule. Clearly, since the growth rate for the energy usually involves derivatives, 
in general there would still be a discrepancy between the analytical and
numerical estimates even if the Leibnitz rule was satisfied, 
arising from the difference between numerical and
analytical derivatives. These differences, however, can also be controlled  
by appropriately rearranging terms (see, for example, Sections \ref{wave_eqs}-\ref{max_eqs}, and
 Refs. \cite{paper0,paper1,gioeldave}).

In the following we discuss more complex examples:  the propagation 
of scalar and electromagnetic fields on the Schwarzschild spacetime 
in three dimensions, and one-dimensional linear perturbations of  
gauge waves. In the former cases, we do have readily available sharp 
energy estimates derived from physical principles. In the latter a 
sharp estimate is available for the subsidiary system describing the 
constraint evolution.

%%%%%%%%%%%%%%%%%%%%%%%%%%%%%%%%%%%%%%%%%%%%%%%%%%%%%%%%%%%%%%%%%%%%%%%% 
\section{Formulation of sample problems} \label{equations} 
%%%%%%%%%%%%%%%%%%%%%%%%%%%%%%%%%%%%%%%%%%%%%%%%%%%%%%%%%%%%%%%%%%%%%%%%%%% 
 
In this section we describe some model problems commonly used in relativity, 
while corresponding numerical results are presented in the 
following section.  The first two problems  
are  Klein--Gordon and Maxwell fields on a Schwarzschild  
black hole background, and linearization perturbations of the gauge wave is the third one. 
 
\subsection{Scalar and Maxwell fields on a black hole background.} 
 
Black holes contain physical curvature 
singularities that can not be included on the computational domain. 
Black hole excision eliminates this problem 
by removing a region around the singularity from the  
computational domain~\cite{unruh}. This region must lie within the  event 
horizon, and be {\it causally disconnected} from the spacetime outside 
the black hole so that the excised region does 
not affect the physics outside the hole.  As discussed in the appendix, 
this region must be carefully chosen for these conditions to hold. 
 
Numerical implementations of excision have traditionally presented a challenge in 
deciding how points near the inner boundary are updated.  Especially in 
higher dimensions, many combinations of one-sided derivative stencils, 
interpolation, and extrapolation have been investigated, mostly through
experimentation ~\cite{cardiff,cornell,aei,uiuc,psu,andersonmatzner} 
(for an alternative excision strategy using pseudospectral methods see \cite{spectralexcision}). 
However, as explained in paper I, derivative operators that satisfy SBP 
together with a diagonal inner product uniquely specify the excision 
algorithm.  The procedure is unambiguous, and for linear problems 
guarantees a stable implementation.  We follow this procedure, detailed 
in Paper I,  to excise the singularity. 
 
We consider three-dimensional evolutions on a Schwarzschild background,  
in either Painlev\' e--Gullstrand (PG) and Kerr-Schild (KS) coordinates.   
The four-dimensional metric, written in terms of ADM quantities, is 
\begin{equation} 
ds^2=-\alpha^2  dt^2 + h_{ij}(dx^i + \beta^i dt)(dx^j+ 
\beta^j dt) \, . 
\end{equation} 
In PG coordinates 
\begin{equation} 
\alpha =1 \;\;\; , \;\;\; \beta^i = \left( \frac{2m}{r} \right)^{1/2} x^i \;\;\; , \;\;\;h_{ij} = \delta_{ij} \, , 
\end{equation} 
while in KS coordinates 
\begin{eqnarray} 
\alpha &=& \left( \frac{r}{r+2M} \right)^{1/2} \; ,\nonumber\\ 
\beta^i &=& \frac{2M}{r+2M} x^i \; ,\nonumber\\ 
h_{ij} &=& \delta_{ij} + \frac{2M}{r} \frac{x^i x^j}{r^2}\; , 
\end{eqnarray} 
where $r=(x^2+y^2+z^2)^{1/2}$ and $M$ is the black hole's mass.  
We excise a cubical excision region centered on the black hole, and  
require the length of the cube, $L$, 
to satisfy $L<4M/(3\sqrt{3})$, or $L \lesssim 0.78M$.  Larger cubes, as  
discussed in the appendix, result in excision boundaries that are {\it not} 
purely outflow.

%%%%%%%%%%%%%%%%%%%%%%%%%%%%%%%%%%%%%%%%%%%%%%%%%%%%%%%%%%%%%%%%%%%%%%%% 
\subsubsection{The Klein--Gordon field on a Schwarzschild background} \label{wave_eqs} 
%%%%%%%%%%%%%%%%%%%%%%%%%%%%%%%%%%%%%%%%%%%%%%%%%%%%%%%%%%%%%%%%%%%%%%%% 
 
The massless Klein--Gordon scalar wave equation,  
$g^{ab}\nabla _a \nabla_b \Phi =0$, 
where $g^{ab}$ is the inverse four-dimensional metric, can be 
be written as
\begin{eqnarray} 
\partial_t \Phi  &=& \Pi, \nonumber\\ 
\partial_t \Pi &=& \beta^i \alpha^{-1}\partial_i(\alpha \Pi) +  
h^{-1/2}\partial_i(h^{1/2}\beta^i\Pi) \nonumber\\ 
       &&  + h^{-1/2}\partial_i(\alpha h^{1/2}H^{ij}d_j) 
  \nonumber\\ 
\partial_t d_i &=& \partial_i \Pi \label{eq:wave1}, 
\end{eqnarray} 
where we have introduced new variables $\Pi := \partial_t \Phi$ and  
$d_i := \partial_i \Phi$ to write the system in first order form. 
Additionally, we define 
$H^{ij}:=h^{ij}-\alpha^{-2} \beta^i \beta^j$, where $h^{ij}$ is the inverse of 
the three-metric, $h_{ij}$, and $h=\det(h_{ij})$. 
 
This formulation has a number of desirable features when the background 
metric is stationary~\cite{paper0}.  For example, Eqs.~(\ref{eq:wave1}) 
are symmetric with respect to the physical energy, and a straightforward 
estimate shows that this energy does not grow when maximally dissipative 
boundary conditions, which are automatically constraint-preserving 
for this formulation, are given.  Moreover, Eqs.~(\ref{eq:wave1}) have 
been written such that the Leibnitz rule is not needed to guarantee that the
energy will not grow.  
For this reason, we refer to the discretization obtained by 
replacing $\partial_i $ by $D_i$ in  Eqs.~(\ref{eq:wave1}) as {\it energy 
preserving} or {\it strictly stable}. The physical energy, however,  is 
not positive definite inside the black hole, since the Killing vector 
becomes space-like there.  We nevertheless 
 choose this energy-preserving discretization, since it guarantees 
long-term stability if the computational domain does not include the black 
hole.  More sophisticated formulations and discretizations, which are
globally symmetric hyperbolic and
conserve the physical energy in the exterior region, up to a distance
arbitrarily close to the event horizon, can also be
constructed~\cite{paper0}.

%%%%%%%%%%%%%%%%%%%%%%%%%%%%%%%%%%%%%%%%%%%%%%%%%%%%%%%%%%%%%%%%%%%%%% 
\subsubsection{Maxwell fields on a Schwarzschild background}  \label{max_eqs}
%%%%%%%%%%%%%%%%%%%%%%%%%%%%%%%%%%%%%%%%%%%%%%%%%%%%%%%%%%%%%%%%%%%%%% 
 
As a second example of classical fields on a black hole background we 
choose the vacuum Maxwell equations. 
We write the Maxwell equations in terms of the Faraday tensor, $F_{ab}$, and 
its dual 
\begin{eqnarray} 
\label{eq:Max} 
\nabla_{[a} F_{bc]} &=& 0 \nonumber \\ 
\nabla_{[a} {}^{\star}F_{bc]} &=& 0. 
\end{eqnarray} 
The dual of the Faraday tensor is 
${}^{\star}F_{ab} := \frac12 \eps_{ab}{}^{cd} F_{cd}$, where $\eps_{abcd}$ 
is the completely anti-symmetric symbol. Contracting  
the above equations with a time-like vector field $u^a$ we get  
the evolution equations: 
\begin{eqnarray} 
\label{eq:Max_evol_1} 
{\cal L}_u F_{ab} &=& - 2 \nabla_{[a}(F_{b]c}u^c),\nonumber \\ 
{\cal L}_u {}^{\star}F_{ab} &=& - 2 \nabla_{[a}({}^{\star}F_{b]c}u^c). 
\end{eqnarray} 
 
Defining a foliation by dragging 
an initial space-like hypersurface along the integral curves of the 
vector $u^a$ (each labeled with the proper time of the  
integral lines of $u^a$, $t$), and  
pulling back the above equations onto each of these hypersurfaces 
we get 
\begin{eqnarray} 
\label{eq:Max_evol_2} 
{\cal L}_u \phi_{\star} (F_{ab}) &=&  
          - 2 \partial_{[a}\phi_{\star}(F_{b]c}u^c),\nonumber \\ 
{\cal L}_u \phi_{\star}({}^{\star}F_{ab}) &=&  
          - 2 \partial_{[a}\phi_{\star}({}^{\star}F_{b]c}u^c), 
\end{eqnarray} 
where $\phi_{\star}$ is the pull-back map into the hypersurface. 
Notice that only exterior derivatives appear in the equations, so 
the geometry only enters through the star operation. 
The electric and magnetic fields are typically defined as 
$E_{a} := F_{ab}n^b$, and $B_a := \frac12 \eps_a{}^{bcd} F_{bc}n_d$,  
allowing one to decompose $F_{ab}$  and ${}^{\star}F_{ab}$ into ($n^a$ 
defined by  $u^a = (\partial_t)^a = \alpha n^a + \beta^a$)
\begin{eqnarray} 
\label{eq:Faraday_decomp} 
F_{ab} &=& -2 E_{[a} n_{b]} + \eps_{ab}{}^{cd} B_c n_d ,\nonumber\\ \,  
{}^{\star}F_{ab} &=& 2 B_{[a} n_{b]} + \eps_{ab}{}^{cd} E_c n_d. 
\end{eqnarray} 
Contracting Eqs.~(\ref{eq:Max_evol_2}) with  
$\frac12 \eps^{abcd}n_d$, 
we get  
\begin{eqnarray} 
{\cal L}_u E_i &=& \eps_i{}^{jk} \partial_j(\alpha B_k - \eps_{klm} \beta^k
E^m) \label{max1} \\ 
{\cal L}_u B_i &=& -\eps_i{}^{jk} \partial_j(\alpha E_k + \eps_{klm} \beta^k
B^m) \label{max2} . 
\end{eqnarray} 
As with the Klein-Gordon equation, this system is symmetric hyperbolic 
outside the event horizon, where $u^a$ is time-like,  
and only strongly hyperbolic inside the event horizon, where  
$u^a$ is {\it space-like}, since the symmetrizer is not positive-definite there.   
 
We supplement the above equations with a no-incoming radiation boundary 
condition. No condition is needed to preserve the constraints, for they 
propagate along $u^a$, as are the outer boundaries. Thus, we set the  
incoming modes to zero at a boundary with normal $m_a$,   
via the projector defined as 
\begin{eqnarray} 
P(E,B) &=& I - \sum_{\lambda^+} e_i \theta^i(E,B) \nonumber \\ 
       &=&  (E_i,B_i) \nonumber \\ 
            & & -  \frac{1}{2}(\hat{E}_i  - \eps_i{}^{jk}\hat{B}_i m_k ,  
            \,\hat{B}_i + \eps_i{}^{jk}\hat{E}_i m_k), \label{projMaxwell} 
\end{eqnarray} 
where $e_i$ is the eigenvector basis, $\theta^i$ is the co-basis, we define 
\begin{eqnarray} 
\hat{E}_i &:=& E_i - (m_i-\beta_i)m^jE_j/(\beta^lm_l+1),\nonumber\\ 
\hat{B}_i &:=& B_i - (m_i-\beta_i)m^jB_j/(\beta^lm_l+1), \nonumber 
\end{eqnarray} 
and where the sum 
extends only over eigenvectors with positive eigenvalues.

As with the scalar field, we ask whether the Maxwell equations can be written  
such that the discrete problem has a sharp energy estimate, giving numerical solutions without ALTI. 
The background spacetime is invariant under time translations, implying a conserved quantity or  energy.   
This energy is the integral  
of $T_{ab}n^au^b$, where $u^a$ is the killing vector field (timelike outside
the black hole) over  
the hypersurface.  For the PG foliation this energy is 
\begin{equation} 
\label{eq:E_K} 
{\cal{E}}_K = \frac12 \int [E^iE_i + B^iB_i + 2E^iB^j\beta^k \eps_{ijk}]\,  
dV. 
\end{equation} 
Just as with the Klein--Gordon equation, this expression is not 
positive-definite inside the horizon, reflecting a property of the 
geometry, not of the equation or coordinates used. 
 
To discretize the above system we use the method of lines and substitute  
in Eqs.~(\ref{max1},\ref{max2}) the partial derivatives with difference
operators satisfying SBP. In this way strict stability with 
respect to the discretized version of the above energy is attained.

\subsection{ Linear gauge wave propagation} \label{eqs_gauge}
 
One simple and common test for numerical implementations of the  
Einstein equations is a gauge wave defined by 
\begin{equation} 
ds^2 = e^{A \sin(\pi (x-t))} (-dt^2 + dx^2) + dy^2 + dz^2 , 
\label{gaugewavemetric} 
\end{equation} 
which describes flat spacetime with a sinusoidal coordinate dependence,  
of amplitude $A$, along the $x$ direction.   The non-trivial ADM variables 
for this metric are 
\begin{eqnarray} 
\hat g_{xx} &=& e^{A \sin(\pi (x-t))}\, , \\ 
\hat K_{xx} &=& \frac{A}{2} \pi \cos\left(\pi \left(x-t \right)\right) 
e^{A/2 \sin\left(\pi \left(x-t\right)\right)} \;,  
\end{eqnarray} 
together with the gauge condition 
\begin{eqnarray} 
\hat \alpha &=& e^{A/2 \sin\left(\pi \left(x-t \right)\right)} \, ,\\ 
\hat \beta^i &=& 0 \, . 
\end{eqnarray} 
We evolve the Einstein equations using the symmetric hyperbolic formulation  
presented in~\cite{sarbach-tiglio} with dynamical lapse and 
the time-harmonic gauge. The formulation is cast in first order form  
by introducing the variables  
${\cal A}_x := \partial_x \alpha/\alpha$ and  $d_{xxx} := \partial_x g_{xx}$.  
Rather than solving the full Einstein equations here, we study 
linear perturbations of a background metric given by Eq.~(\ref{gaugewavemetric}).  
Furthermore, we simplify the system by allowing perturbations depending   
only on $(t,x)$, in the form 
\begin{eqnarray} 
g_{xx} &=& \hat g_{xx} + \delta g_{xx}  , \nonumber  \\  
K_{xx} &=& \hat K_{xx} + \delta K_{xx}  , \nonumber \\ 
d_{xxx} &=& \hat d_{xxx} + \delta d_{xxx}  , \nonumber \\ 
\alpha &=& \hat \alpha + \delta \alpha  , \nonumber \\ 
{\cal A}_{x} &=& \hat {\cal A}_{x} + \delta {\cal A}_{x}  . \nonumber 
\end{eqnarray} 
 This restriction can be  justified by noting that 
full non-linear evolutions of data defined by Eq.~(\ref{gaugewavemetric}) 
show that only these variables vary dynamically~\cite{hyperGR}, 
and variations in all other variables are consistent with round-off error.   
The resulting linearized equations are (henceforth dropping the $\delta$ notation) 
\begin{eqnarray} 
\partial_t  \alpha &=& - A \pi \cos(\phi)  \alpha -  K_{xx}  
   + \frac{A \pi}{2 \hat{\alpha}} \cos(\phi)  g_{xx}\, , \label{gauge1}\\ 
\partial_t  {\cal A}_x  &=&  -\frac{1}{\hat{\alpha}} \partial_x  K_{xx}  
       + \frac{A\pi}{2\hat{\alpha}} \cos(\phi)  K_{xx}\, , \nonumber \\ 
   & & -\frac{A\pi^2}{2\hat{\alpha}^2} \left ( A\cos(\phi)^2  
       + \sin(\phi) \right)  g_{xx}   \nonumber \\ 
   & & +\frac{A\pi}{2\hat{\alpha}^2} \cos(\phi)  d_{xxx}    \nonumber \\ 
   & &-\frac{A\pi}{2} \cos(\phi)  {\cal A}_{x} 
       + \frac{A\pi^2}{2\hat{\alpha}} \sin(\phi)  \alpha  \label{gauge2}\, ,\\ 
\partial_t  g_{xx} &=& - A \hat{\alpha} \pi \cos(\phi)  \alpha  
       - 2 \hat{\alpha}  K_{xx} \label{gauge3}\, , \\ 
\partial_t  K_{xx} &=& - \hat{\alpha} \partial_x  {\cal A}_{x}   
       - \frac{A \hat{\alpha} \pi}{2} \cos(\phi)  {\cal A}_{x} \nonumber \\ 
  &  & - \frac{A \pi^2}{4} \left ( -2 \sin(\phi)  
       + A \cos(\phi)^2 \right )  \alpha  \nonumber \\ 
  & &- A \pi \cos(\phi)  K_{xx} \nonumber \\ 
  & & + \frac{A\pi}{4\hat{\alpha}} \cos(\phi)  d_{xxx}\, , \label{gauge4} \\ 
\partial_t  d_{xxx} &=& {A \hat{\alpha} \pi^2} \left ( \sin(\phi)  
      - A \cos(\phi)^2 \right ) \alpha  \nonumber \\ 
  & & - A \hat{\alpha} \pi \cos(\phi)  K_{xx} 
      - A \hat{\alpha}^2 \pi \cos(\phi)  {\cal A}_{x}   \nonumber \\ 
  & & - 2 \hat{\alpha} \partial_x  K_{xx} \, ,\label{gauge5} 
\end{eqnarray} 
where we have defined  $\phi := \pi (x-t)$. 
This system is symmetric hyperbolic; three characteristic speeds 
are $0$ and the other two $\pm 1$.  
 
We consider here an initial boundary value problem on  
the domain $x\in[-1/2,3/2]$.  We implement boundary conditions via the 
orthogonal projector method described in appendix D of Paper~I. 
The positive/negative eigenmode decomposition of the matrix $H A$ is 
$W_{\pm}= - ( {\cal A}_x \pm  K_{xx})$ (see Paper~I). 
Here, the matrix $A$ is the principal part, and $H$ is its  
symmetrizer.  For one-dimensional problems, $H=(P^{-1})^T 
P^{-1}$,  and $P$ is the matrix that diagonalizes $A$, i.e., $P^{-1} A P 
= \Lambda$, with $\Lambda$ a diagonal matrix. 
 
The system must satisfy two constraint equations, corresponding to the  
definition of the variables $d_{xxx}$ and ${\cal A}_x$. When linearized, these 
constraints are 
\begin{eqnarray} 
0 &=& C_x =  - \partial_x  g_{xx} +  d_{xxx} \;, \\ 
0 &=& C_{\cal A} =  {\cal A}_x - \frac{1}{\hat{\alpha }}  
\left ( \partial_x \alpha - \frac{A \pi}{2} \cos(\phi)  
\alpha \right ). 
\end{eqnarray} 
The other constraints are automatically satisfied owing to the restricted 
form of the allowed perturbations.  Despite the simplicity of this system, 
the numerical integration of these equations reveals some interesting 
features. Indeed, straightforward discretizations of this system have 
lead to exponentially growing solutions with quickly growing constraint 
violations.  That this instability has been observed by different 
groups in a variety of situations, led to the speculation that 
constraint violations {\em drive} the instability.  
We offer here a different 
interpretation of the results by examining the evolution of the  
constraints. 
 
Evolution equations for the constraints can be obtained  
by taking the time derivative of $C_x$ and $C_{\cal A}$,  
and substituting in the Einstein equations, giving 
\begin{eqnarray} 
\partial_t C_x &=& -A \pi \hat{\alpha }^2 \cos(\phi) C_A \;, \\ 
\partial_t C_{\cal A} &=& \frac{A \pi}{2\hat{\alpha }^2} \cos(\phi) C_{x} - 
\frac{A\pi}{2} \cos(\phi) C_{\cal A} \; . 
\end{eqnarray} 
These equations can be integrated in closed form to yield  
\begin{eqnarray} 
C_x &=& G(x) \hat{\alpha }^{2} + H(x) \hat{\alpha} \; ,\\ 
C_{\cal A} &=& G(x) + \frac{H(x)}{2\hat{\alpha }} \;, 
\end{eqnarray} 
where $G(x)$ and $H(x)$ are free functions determined by the initial data: 
\begin{eqnarray*} 
G(x)  \hat{\alpha}^2 &=&  2 \hat{\alpha}^2 C_{\cal A}(0,x) - C_x(0,x),  \\ 
H(x) \hat{\alpha} &=& 2 \left( C_x(0,x)  
           - \hat{\alpha}^2 C_{\cal A}(0,x) \right) . 
\end{eqnarray*} 
Clearly the constraint behavior is  {\em purely 
oscillatory}. Hence, any growth in the constraints observed during evolution  
must be purely numerical. We will show in the next section that this is  
indeed the case, and that the error growth can be eliminated.

%%%%%%%%%%%%%%%%%%%%%%%%%%%%%%%%%%%%%%%%%%%%%%%%%%%%%% 
\section{Numerical simulations} \label{simulations} 
%%%%%%%%%%%%%%%%%%%%%%%%%%%%%%%%%%%%%%%%%%%%%%%%%%%%%% 
 
We now present some numerical simulations of the model problems 
discussed in the previous section. For 
the black hole tests, we examine configurations where the computational 
domain is outside the black hole and others that do contain an 
excised black hole. The former configuration allows us to investigate 
the evolution of a symmetric hyperbolic system  with boundary interactions 
without the difficulties introduced by excision.  We then 
examine the more demanding case in which excision is required, 
and the evolution systems here considered are only strongly hyperbolic inside the 
black hole.

%%%%%%%%%%%%%%%%%%%%%%%%%%%%%%%%%%%%%%%%%% 
\subsection{Wave propagation on a Schwarzschild background in  
 KS coordinates} 
%%%%%%%%%%%%%%%%%%%%%%%%%%%%%%%%%%%%%%%%%% 
 
This section presents numerical results for the massless Klein--Gordon 
scalar field on a Schwarzschild black hole background.  We present 
some tests of our code, and then discuss differences between the strictly 
stable and non-strictly stable discretizations.   
 
%%%%%%%%%%%%%%%%%%%%%%%%%%%%%%%%%%%%%%%%%%%%%%%%%%% 
\subsubsection{Case 1: Computational domain outside the black hole} 
%%%%%%%%%%%%%%%%%%%%%%%%%%%%%%%%%%%%%%%%%%%%%%%%%%%%% 
 
Consider an uniform computational domain that  
{\it does not} contain the black hole, $(x,y,z) \in [1.5M, 5.5M]^3$.  
Initial data with compact support are given  
to $\Pi$, the time derivative of the scalar field, as 
\begin{eqnarray} 
\Pi &=& 10^7\left[(x-3M)(x-4M)(y-3M)(y-4M)\right. \nonumber\\ 
 & &\qquad\times \left. (z-3M)(z-4M)\right]^4 
\label{eq:initial_data_1} 
\end{eqnarray} 
when $(x,y,z)\in [3M,4M]^3$, and zero otherwise.  All other fields 
are initially set to zero \footnote{Note that the initial maximum amplitude 
of $\Pi$ is of order one despite the large factor in its definition, 
$\mbox{max}\{\Pi\} = 1.25^7/8 \approx 0.6$.}. This data is then evolved 
using two possible discretizations. The first one is that one obtained 
by direct substitution of  partial derivatives by discrete   
ones in Eq.~(\ref{eq:wave1}), this is the strictly 
stable discretization. The second one is  obtained  
by expanding all derivatives of products with the product rule and then 
replacing partial derivatives of field variables by their discrete 
counterparts. We refer to this as the ``na\" ive'' discretization. 
 
Both discretizations were used to evolve for around 250 crossing times, 
under different choices of Courant factor ($\lambda = 1/4, 1/2, 1$). In all cases we
found no sign of long term 
growth when monitoring the variables, though a closer inspection
reveals several salient features of the obtained solutions. Namely, 
a self-convergence analysis show results consistent with second order
accuracy for a few crossing times (with better values for the smallest
$\lambda$ ). However, at later times, the observed convergence rate deteriorates 
considerably and achieve artificially large values. The time at which this
behavior starts coincides with the fields mostly leaving the computational 
domain. 

\begin{figure}[ht] 
\begin{center} 
\includegraphics*[height=6.cm,width=8.5cm]{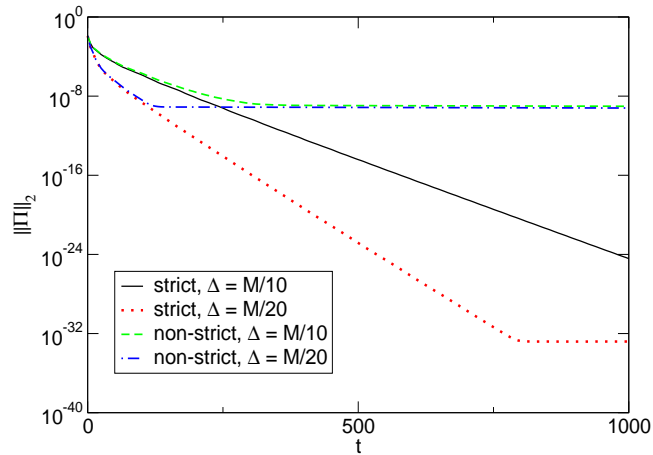} 
\caption{This figure shows the $L_2$ norm of $\Pi$ in a scalar field 
evolution outside the black hole for about 250 crossing 
times.  The figure compares 
results for a na\"ive discretization (see text), and a strictly 
stable discretization at two resolutions, $\Delta=M/10$ and $\Delta=M/20$. 
The domain is $(x,y,z)\in[1.5,5.5]$, $\lambda=0.5$, and no dissipation 
is added to the solution. 
} 
\label{conserve_straight_outside}  
\end{center} 
\end{figure}

For instance,  Figure \ref{conserve_straight_outside} displays the 
$L_2$ norm of $\Pi$ obtained with  the strictly stable and ``na\" ive'' 
discretizations. 
The figure shows the $L_2$ norms of $\Pi$ calculated at two resolutions,  
for 250 crossing times.  The large decrease in the norm is a sign that basically
most of the fields leave the domain after a few crossing times. 
 Both discretizations give a long-term stable  
algorithm for this case.  But, as we will discuss later,    
the na\"ive discretization without dissipation becomes long-term unstable 
when the black hole is included on the domain.
 
\begin{figure}[ht] 
\begin{center} 
\includegraphics*[height=6cm]{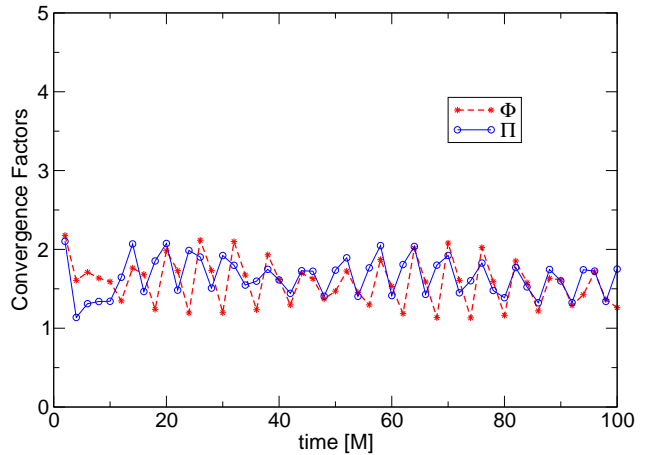} 
\caption{ 
This figure shows the self-convergence factor as a function of time  
for the fields $\Pi$ (solid line) and $\Phi$ (dashed line) 
on a domain outside the black hole, using the strictly stable scheme (similar
factors are obtained for the naive one), for about $25$ crossing times.   
These factors are calculated using resolutions of $M/10$, $M/20$, and $M/40$. 
The obtained factors are close to the expected value of two.
The Courant factor is $\lambda=1$, and no dissipation is used. } 
\label{long_conv}  
\end{center} 
\end{figure}

To verify that we get the expected self-convergence factor in long term 
simulations when the fields do not decay to very small values after having 
left the domain, we now show results of a complementary set of tests on the 
domain $x\in[2.1M,6.1M]$, $(y,z)\in[-2M,2M]$. Here we concentrate in what
corresponded to the worst case scenario considered, given by $\lambda = 1$. 
We set the incoming modes on all outer faces to zero, except on the 
$x=6.1 M$ face,   where the time derivative of the incoming fields is set to 

\begin{eqnarray} 
W_{+} = \left\{ \begin{array}{ll} 
        (y^2+z^2-M)^6 (M/t-1)^6 \sin(t)& \nonumber\\ 
        \qquad\qquad\qquad\mbox{if $t\ge1$,}  
                  \mbox{$(y,z)\in[-M,M]$}& \nonumber \\ 
        0\qquad\qquad\qquad\mbox{otherwise} &{} \nonumber 
        \end{array} 
        \right. \label{eq:incoming_modes} 
\end{eqnarray}  
This boundary condition is imposed through an 
orthogonal projection, as explained in Paper I.
Initial data is given by 
\begin{eqnarray} 
\Pi &=& 10^7\left[(x-3M)(x-4M)(y-M)(y+M)\right. \nonumber\\ 
 & &\qquad\times \left. (z-M)(z+M)\right]^4 \;,  
\label{eq:initial_data_1b} 
\end{eqnarray} 
and both discretizations (strictly and non-strictly stable) are evolved.   
Figure~\ref{long_conv} shows the self-convergence factor (as before, in the $L_2$ norm)
 of $\Phi$ and $\Pi$ versus time for the strictly stable discretization (similar factors are
obtained for the non-strictly stable one) at the resolutions $\triangle =M/10$, 
$M/20$ and $M/40$. Convergence factors close  
to the expected value of two are observed, with no qualitative difference in the convergence 
factor as time progresses. 
 
%%%%%%%%%%%%%%%%%%%%%%%%%%%%%%%%%%%%%%% 
\subsubsection{Case 2: Domain with excision} 
%%%%%%%%%%%%%%%%%%%%%%%%%%%%%%%%%%% 
 
In this section,  the computational domain contains the black hole, which 
is excised with an inner cubic boundary centered on the black hole.  The 
domain is defined on $[-4.5M, 4.5M]^3$, and the excision cube  
has a total length of $0.6M$.  Thus, the faces of the inner boundary  
are at $\pm0.3M$.   
 
\begin{figure}[ht] 
\begin{center} 
\includegraphics*[height=6cm]{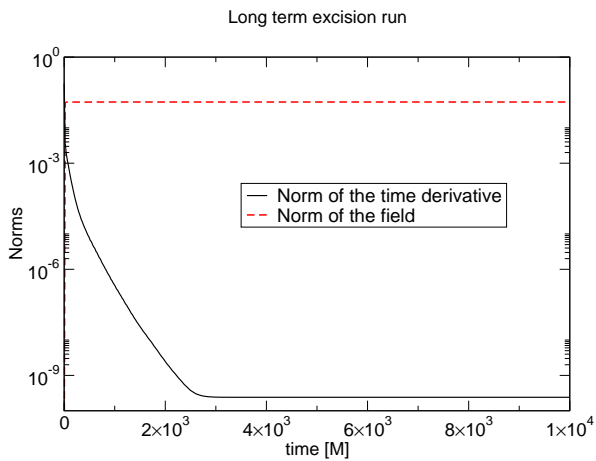} 
\caption{ 
This figure demonstrates the long-term stability of the strictly 
stable discretization of the scalar wave equation when an excised 
black hole is on the grid.  The $L_2$ norms of $\Pi$ (solid line) 
and $\Phi$ (dashed line) are shown for about 2500 crossing times. 
No slowly growing unstable modes are detected in these runs. 
This run was performed with $81^3$ grid points, $\lambda=0.25$, 
and no dissipation is used. 
} 
\label{long_excision}  
\end{center} 
\end{figure}

\begin{figure}[ht] 
\begin{center} 
\includegraphics*[height=6cm]{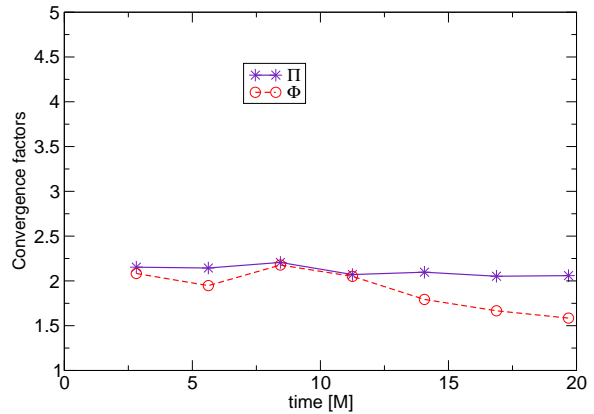} 
\caption{ 
This figure shows a self-convergence test for $\Pi$ (solid line) 
and $\Phi$ (dashed line) using the same  
configuration as in Fig.~\ref{long_excision}.   The self-convergence 
test is performed on uniform grids with $81^3, 161^3$, and $321^3$ grid-points.  
The convergence factor is close to the expected value of two.  The measured convergence 
factor drops slightly from near two after one crossing time, when most  
of the field has left the domain and afterwards increases to unrealistic 
high values. 
} 
\label{convergence_excision}  
\end{center} 
\end{figure}

\begin{figure}[ht]
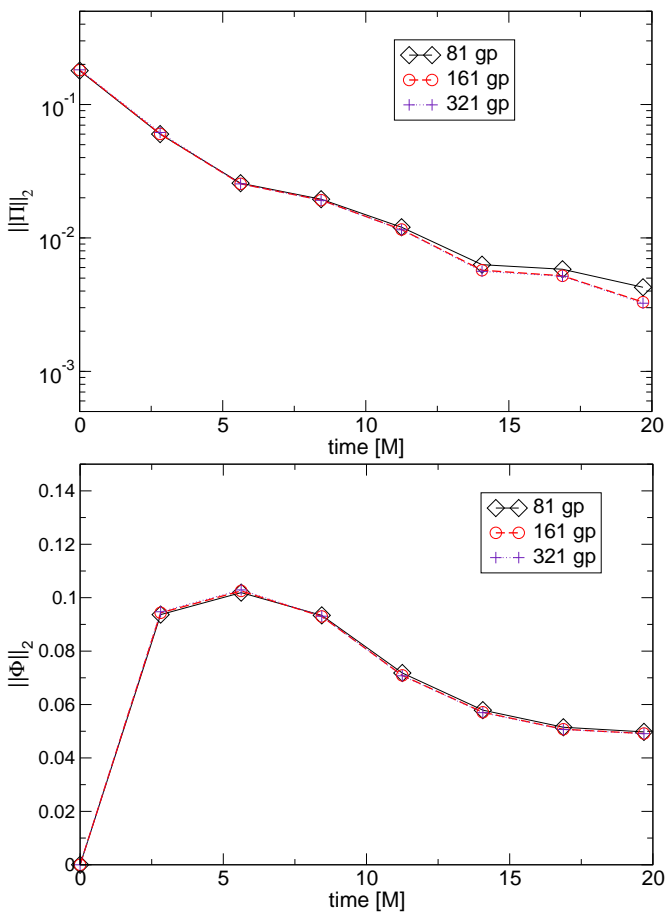
 
\begin{center} 
\includegraphics*[height=6cm]{exc_d4phi_norms.eps} 
\includegraphics*[height=6cm]{exc_phi_norms.eps} 
\caption{ 
This figure shows 
$L_2$ norms for the  runs shown in Fig.\ref{convergence_excision}. The 
relative difference in the norms for $\Phi$ at any 
two different resolutions, is at most about one percent. Similarly for $\Pi$, 
except at the last point shown, where this field has mostly left the grid. 
} 
\label{norm_excision}  
\end{center} 
\end{figure}

We first verify that the excision algorithm does not display ALTI when 
employing the strictly stable scheme.  As mentioned 
earlier, very little freedom is available in constructing difference 
operators that satisfy SBP.  Indeed, for our second-order code with 
a diagonal scalar product, the stencils are uniquely specified.   
Figure~\ref{long_excision} shows the $L_2$ norms 
of $\Pi$ and $\Phi$ for a long --up to $10^4M$-- evolution (with no dissipation), 
designed to detect slowly 
growing modes.  As shown in this figure, the algorithm specified by 
requiring SBP and the equations arranged so as to preserve the energy is
long-term stable even 
with excision boundaries.  For this run non-trivial initial data are given, as
before, only to $\Pi$. 
\begin{eqnarray} 
\Pi = \left\{ \begin{array}{ll} 
        \delta^{-8} r^{-1}\left[ (r-r_0)^2-\delta^2 \right]^3 (r0-r) &
        \mbox{if $|r-r_0|\le |\delta |$}  \\ 
        0 & \mbox{otherwise \, .}  
               \end{array} 
        \right.  
\end{eqnarray}  
with $r_0=2.5M$ and $\delta = 1.5M$. When studying the convergence of the 
code one again faces the fact that after a few crossing times the convergence 
value become meaningless. The behavior observed is analogous to the case where 
the black hole was outside the computational domain. Namely, close to second order convergence 
is seen during the first few crossing times, and unrealistically high values are obtained 
afterwards. In Fig.~\ref{convergence_excision} we show a short-term convergence 
test (about two crossing times) with excision.   Fig.~\ref{norm_excision} 
shows the corresponding $L_2$ norms for $\Phi$ and $\Pi$.

\begin{figure}[ht] 
\begin{center} 
\includegraphics*[height=6cm]{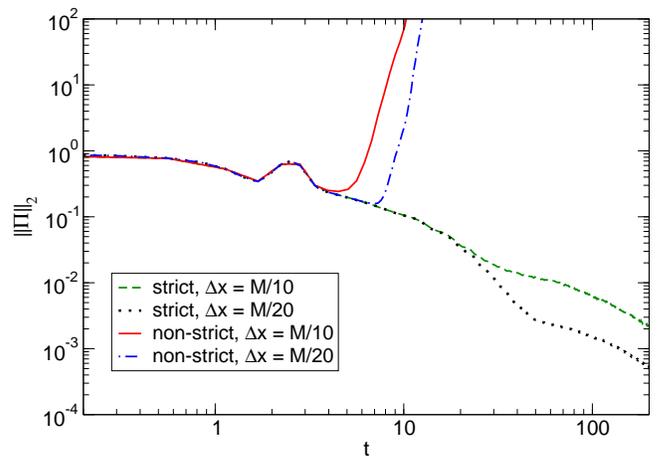} 
\caption{ 
This figure shows the $L_2$ norm of $\Pi$ in a scalar field 
evolution on a domain containing an excised Schwarzschild black hole. 
The solid and dot-dashed lines show the $L_2$ norm of $\Pi$ obtained 
using the na\"ive discretization of the wave equation.  Although the 
amplitude decrease with resolution, this solution quickly becomes long-term unstable, 
indicating an artificial instability. 
The strictly stable discretization (shown with dashed and dotted lines),  
however, remains long term stable and well-behaved.  
The results are calculated with the resolutions 
$\Delta=M/9$ and $\Delta=M/18$, $\lambda=0.5$, and no dissipation  
is added. 
} 
\label{conserve_straight_inside}  
\end{center} 
\end{figure}

\begin{figure}[ht] 
\begin{center} 
\includegraphics*[height=6cm]{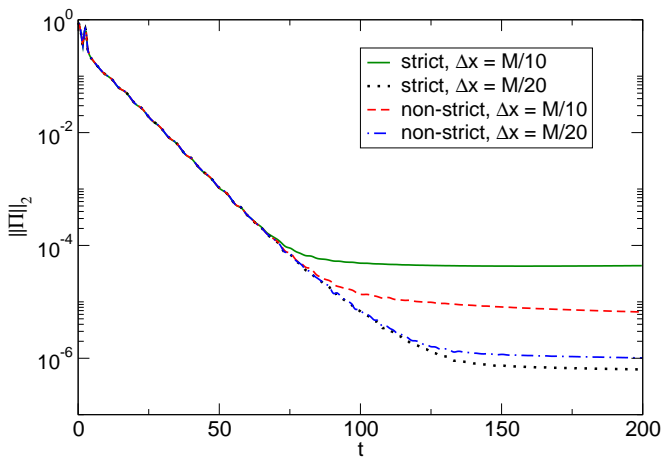} 
\caption{ 
This figure shows the $L_2$ norm of $\Pi$ in a scalar field 
evolution on a domain containing an excised Schwarzschild black hole. 
The solid and dot-dashed lines show the solutions obtained 
using the na\"ive discretization with dissipation, the latter eliminating the  
artificial instability noted in Fig.~\ref{conserve_straight_inside}. 
The strictly stable discretization (shown with dashed and dotted lines), 
without dissipation, is shown for comparison. 
The results are calculated with the resolutions 
$\Delta=M/9$ and $\Delta=M/18$, $\lambda=0.5$, and the dissipation 
parameter is $\sigma=0.02$. 
} 
\label{conserve_straight_dissip}  
\end{center} 
\end{figure}

As in the previous case, we also compare results obtained with  
the strictly stable and the na\"ive discretizations. 
 Long term stability was found for both cases  
when the computational domain was outside the black hole. 
With excision, however, the solution obtained with the  
na\"ive discretization quickly becomes  
corrupted, as shown in Fig.~\ref{conserve_straight_inside}.  As 
shown in this figure, the amplitude decreases with resolution, indicating 
that the growth is spurious and that the na\"ive discretization suffers from ALTI.   The strictly 
stable discretization, however, with a sharp semi-discrete energy estimate, 
remains well-behaved.  The ALTI that contaminate the na\"ively discretized 
solution are found to be of high-frequency type, which can be managed by adding 
explicit dissipation.  Figure~\ref{conserve_straight_dissip} shows 
results using the same initial configurations, except that now  
 dissipation is added to the na\"ive discretization. 
Now all of the runs produce long term stable simulations.

%%%%%%%%%%%%%%%%%%%%%%%%%%%%%%%%%%%%%%%%%% 
\subsection{Maxwell fields in a Schwarzschild background in  
PG coordinates} 
%%%%%%%%%%%%%%%%%%%%%%%%%%%%%%%%%%%%%%%%%% 
 
As with the scalar field tests above, we perform tests with the 
Maxwell equations on two kind of domains.  Specifically, we choose 
 
\begin{enumerate} 
\item Computational domain outside the black hole:  
   $ y,z\in[-4M,4M]^2$ and $x\in [3M,11M]$;  
\item Computational domain including part of the black hole: $(x,y,z) \in
  [-8.5M,8.5M]^3$. The singularity is excised by means of a cubical  
   region defined by $(x,y,z) \in [-0.35M,0.35M]^3$.  
\end{enumerate} 
The initial data describe a ``pulse'' of electromagnetic field given by 
\begin{eqnarray*} 
F_i = \left\{ \begin{array}{ll} 
        \alpha\delta^{-8}\left[(r-r_0)^2-\delta^2 \right ]^8
        \epsilon_{ij3} (x^j-x_o^j) & \mbox{if $|r-r_0| < \delta$}  \\ 
        0 & \mbox{otherwise \, ,}  
               \end{array} 
        \right.  
\end{eqnarray*}  
where $F_i$ stands for $E_i$ and $B_i$, $\alpha$ is a strength 
parameter,  
$x_o^i$ is the location of the pulse's center,  and $\delta$ its width. 
We discretize in a strictly stable way, as explained in Section \ref{max_eqs}. 
 
\subsubsection{Case 1 (domain outside the black hole)} 
 
\begin{figure}[ht] 
\begin{center} 
\includegraphics*[height=6cm]{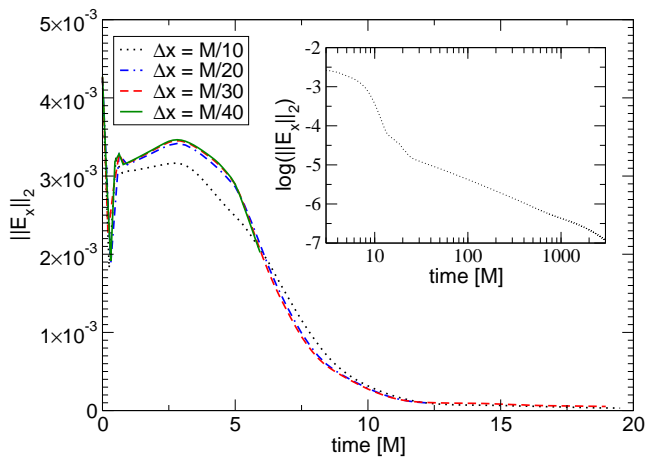} 
\caption{This figure shows the $L_2$ norm of $E_x$ vs.\  time in 
an evolution of the Maxwell fields at  
four different resolutions for a strictly stable discretization. The computational  
domain is outside the 
black hole. The inset shows a long-term evolution, $3000M$, using the 
coarsest resolution.  The runs were performed on uniform grids 
with resolutions $M/10$, $M/20$, $M/30$ and $M/40$, and the Courant factor is $\lambda =0.25$. 
} 
\label{maxwell12}  
\end{center} 
\end{figure}

\begin{figure}[ht] 
\begin{center} 
\includegraphics*[height=6cm]{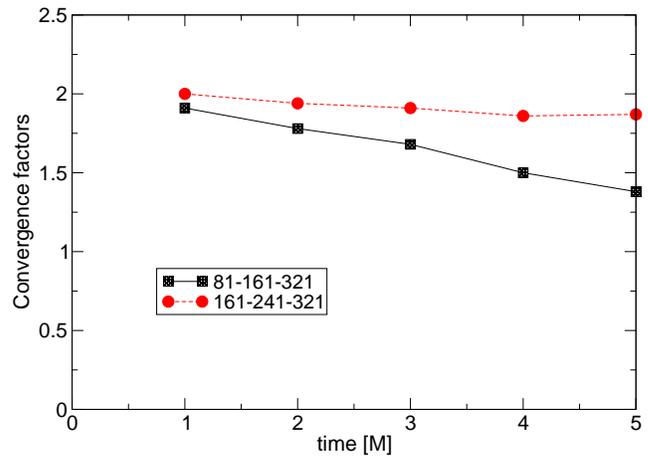} 
\caption{ 
This figure shows self-convergence factors for $E_x$ for the runs of  
Fig.~\ref{maxwell12}. The factors improve with resolution, in the sense that
they closer to the expected value of two.  The labels $81$-$161$-$321$ and $161$-$241$-$321$  
indicate the number of grid points in 
each direction used to perform the self-convergence test. 
} 
\label{maxwell11}  
\end{center} 
\end{figure}

Let $\alpha=1$, $x_o^i=(7M,0,0)$ and $\delta=M$, and we set the 
dissipation to zero. Figure~\ref{maxwell12} shows $||E_x||_2$ vs.\  
time for four different resolutions 
($81^3,161^3,241^3$ and $321^3$ grid-points, corresponding to $\Delta = 
M/10,M/20,M/30$ and $M/40$, respectively),  while the inset shows the 
decay in time for a long-term run with the coarsest resolution.  
Figure~\ref{maxwell11} shows the self-convergence factor of $E_x$ 
for a short time, computed with $81^3$, $161^3$ and $321^3$ grid-points,  
and $161^3$, $241^3$, and $321^3$ 
grid-points. As expected, the factor gets closer to two when resolution 
is increased. 
 
\subsubsection{Case 2: Domain with excised black hole} 
 
\begin{figure}[ht] 
\begin{center} 
\includegraphics*[height=6cm]{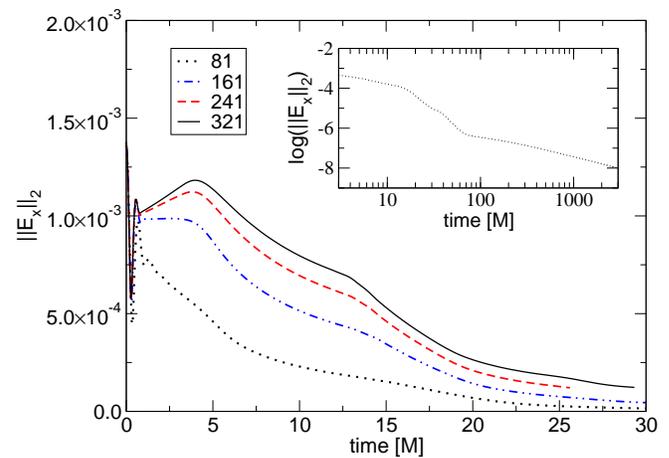} 
\caption{ 
This figure shows the $L_2$ norm of $E_x$ vs.\  time in 
an evolution of the Maxwell fields on a computational domain containing 
an excised black hole for a strictly stable discretization.  The norms are 
 shown for four different resolutions,  
and the inset shows a long-term evolution, $3000M$, at the 
coarsest resolution.  The runs were performed on uniform grids 
with resolutions $M/10$, $M/20$, $M/30$ and $M/40$.  The Courant factor 
is $\lambda=0.25$, and dissipation is added to the evolution equations, with  
$\sigma=0.05$. 
} 
\label{maxwell22} 
\end{center} 
\end{figure}

\begin{figure}[ht] 
\begin{center} 
\includegraphics*[height=6cm]{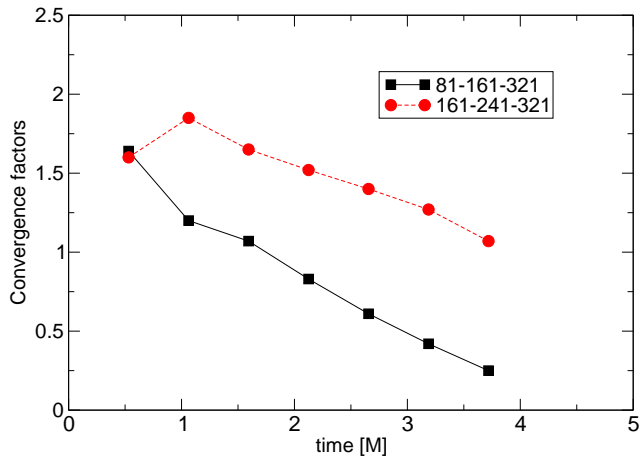} 
\caption{ 
This figure shows convergence data for the runs of 
Fig.~\ref{maxwell22}. The self-convergence factors are calculated from 
$E_x$.  The excision 
boundary is very close to the black hole, and the large gradients 
encountered in the fields and spacetime near the excision boundary 
require very high resolution to resolve them adequately.  The coarsest 
solutions are not resolved well enough to be in the convergence
regime. However, it can be seen from the Figure that the convergence improves
with resolution.  
The labels $81$-$161$-$321$ and $161$-$241$-$321$ 
indicate the number of grid points in 
each direction used to perform the self-convergence test. 
} 
\label{maxwell21}  
\end{center} 
\end{figure}

Now let $\alpha=1$, $x_o^i=(6.5M,0,0)$ and $\delta = M$. In contrast 
to the scalar field runs, solutions of the Maxwell equations without  
explicit dissipation diverge at long times, even with a strictly stable  
discretization. The reason for this appears to be that the energy inside the
black hole 
is not positive definite and boundedness (either at the continuum or numerical
level) is not guaranteed (it is
not clear why this divergence does not appear in the scalar field case, when
the singularity is excised, since the energy is also non-positive definite in
that case.).  
High frequencies are seen in the solution,  
and we therefore add a small amount of dissipation, $\sigma=0.05$. 
Figures~\ref{maxwell22} and \ref{maxwell21} show the norms and convergence 
data similar to the figures for the domain outside the black hole. 
As before, the simulations are long-term stable for long times.  The self-convergence 
factors for the coarsest resolutions drop quickly from their initial value, 
indicating that the coarsest solutions are not in the convergent regime. 
In addition to effects caused by the fields leaving the computational 
domain, 
using a cubical excision boundary forces us to excise very close to the 
black hole, where the spacetime and field gradients are very large, effectively 
requiring higher resolution.

%%%%%%%%%%%%%%%%%%%%%%%%%%%%%%%%%%%%%%%%%% 
\subsection{Linearization around the gauge wave} 
%%%%%%%%%%%%%%%%%%%%%%%%%%%%%%%%%%%%%%%%%% 
 
The gauge wave is a useful test problem for examining 
some of the difficulties found in the simulation of the Einstein 
equations (see ~\cite{convergence,gaugeapples,gaugejansen,winicourgauge}),  
and to illustrate the usefulness of 
the techniques presented in Paper I. To this end, we present results 
obtained using a straight-forward discretization of the right hand side 
of Eqs.~(\ref{gauge1})--(\ref{gauge5}), 
and show results obtained for long term evolutions with and without 
 dissipation.  Moreover, we consider  cases both where 
the constraints are initially satisfied, and where they are not.  
The latter illustrates 
that even when large constraint violations are introduced in the  
initial data, the solutions to the linearized equations remain well behaved. 
This, of course, may not necessarily be
the case for the non-linear Einstein equations, as for these relatively large amplitudes 
 the linearized system need not be a close approximation to the non-linear ones. Or
 even if the amplitudes were small, there could be purely non-linear effects
 in the solution which would naturally be missed in a linearized analysis \cite{sartom}.

For these tests we adopt trivial initial data for all fields, except for 
$g_{xx}$,  $d_{xxx}$ and $K_{xx}$, which have compact support in $x \in 
[0,1]$ and are  given by 
\begin{eqnarray} 
 g_{xx} &=& \kappa \sin(x) \left [ (x-0.5)^2 - 0.25 \right ]^4, \\ 
 d_{xxx} &=& \mu \partial_x  g_{xx}, \\ 
 K_{xx} &=&  -\frac{\xi}{2} \partial_x  g_{xx} \, , 
\end{eqnarray} 
where $\kappa$, $\mu$, and $\xi$ are constants.  We set $A=0.1$ in the 
background solution and the computational 
domain is taken to be $x\in[-0.5,1.5]$. Boundary conditions 
are applied via Olsson's projection operators. Here, as we are interested 
in the long term behavior of the solution, we couple the incoming modes 
the outgoing ones such that 
$$
W_+=0.5 W_- \; .
$$  
For cases where the constraints are initially satisfied, we choose 
$\mu=1=\xi$, $\kappa=0.01$; otherwise we set $\mu=0$, 
$\kappa=0.01$ and $\xi=1$. In all these simulations a Courant factor of
$\lambda=0.25$ is used, to compare with similar simulations   
 ~\cite{convergence,gaugeapples,gaugejansen,winicourgauge}.

\begin{figure}[ht] 
\begin{center} 
\includegraphics*[height=6cm]{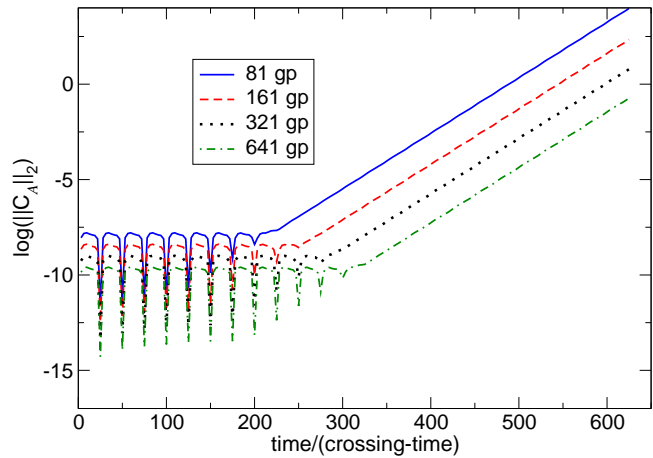} 
\caption{Logarithm of the $L_2$ norm of $C_{{\cal A}}$ vs. number of crossing  
times for four different resolutions. After 
an initial stage where the constraints are well behaved a clear exponential 
mode dominates the solution.}. 
\label{gaugewave_C_straight}  
\end{center} 
\end{figure}

\begin{figure}[ht] 
\begin{center} 
\includegraphics*[height=6cm]{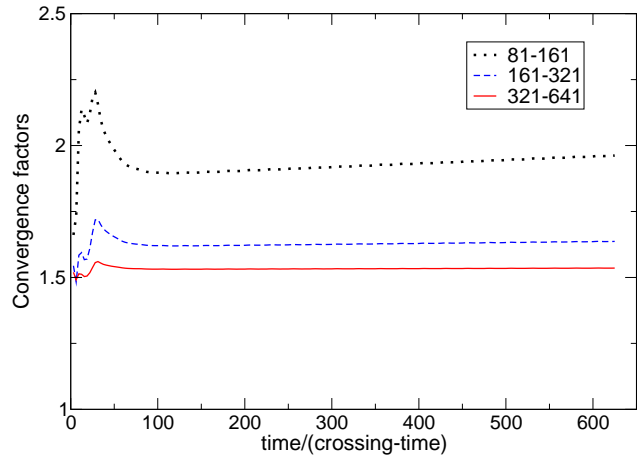} 
\caption{Convergence factors (to zero) for $C_{{\cal A}}$ as a function of  number of  
crossing times, comparing runs with $81$ and $161$ points (dotted line), 
$161$ and $321$ points (dashed line) and $321$ and $641$ points (solid line).  
Notice that even when the solutions 
are growing exponentially (see Figs.\ref{gaugewave_C_straight},\ref{gaugewave_g_straight}),  
the constraints converge to zero.}. 
\label{gaugewave_CON_straight}  
\end{center} 
\end{figure}

\begin{figure}[ht] 
\begin{center} 
\includegraphics*[height=6cm]{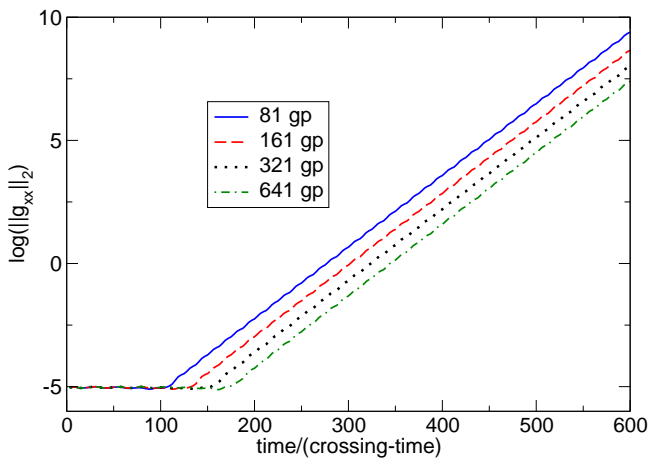} 
\caption{Logarithm of the $L_2$ norm of $g_{xx}$ vs. number of crossing  
times for four different resolutions with no dissipation.  
At early  times, the norm remains fairly constant 
but at later ones an exponential growth dominates the solution.}. 
\label{gaugewave_g_straight}  
\end{center} 
\end{figure}

Figures \ref{gaugewave_C_straight}-\ref{gaugewave_g_straight} show results
for constraint-satisfying initial data, without dissipation. 
Figure~\ref{gaugewave_C_straight} shows the $L_2$ norm 
of the constraint, $C_{{\cal A}}$ and its convergence to zero with increasing
resolution. 
Initially  the norm $C_{{\cal A}}$ is 
almost constant.  However, at later times a 
clear exponential mode is observed.  While the solution is well-behaved 
for longer times with increasing resolution, this approach is 
impractical for long runs. Figure~\ref{gaugewave_CON_straight} shows the convergence (to zero)  
factors in the $L_2$ norm obtained for $C_{{\cal A}}$. For the highest
resolutions, the factor is very close to its expected value.  

An exponential behavior for $g_{xx}$ is also seen in Figure
\ref{gaugewave_g_straight}. 
Even though in principle this quantity could already grow at the
continuum, the fact that at fixed time the norm of $g_{xx}$ decreases 
with increasing resolution  suggests that, as discussed below, this is
 an artifact of an ALTI.

\begin{figure}[ht] 
\begin{center} 
\includegraphics*[height=6cm]{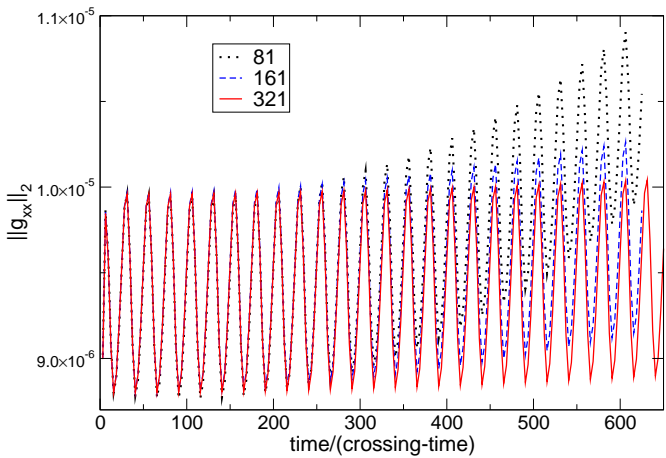} 
\caption{$L_2$ norm of $g_{xx}$ vs. number of crossing times for four  
different resolutions.  Only a small growth is observed in the 
solution.}. 
\label{gaugewavediss_g_straight}  
\end{center} 
\end{figure}

\begin{figure}[ht] 
\begin{center} 
\includegraphics*[height=6cm]{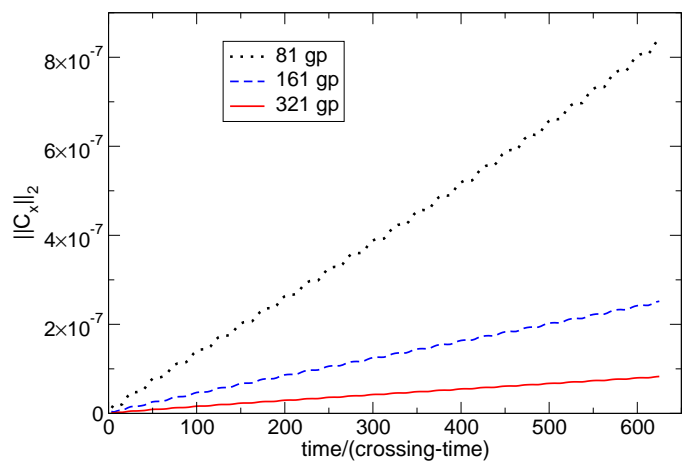} 
\caption{$L_2$ norm of $C_{{\cal A}}$ vs. number of crossing times for four  
different resolutions. Convergence to zero is observed and only slow (linear)   
growth is observed.}. 
\label{gaugewavediss_C_straight}  
\end{center} 
\end{figure}

\begin{figure}[ht] 
\begin{center} 
\includegraphics*[height=6cm]{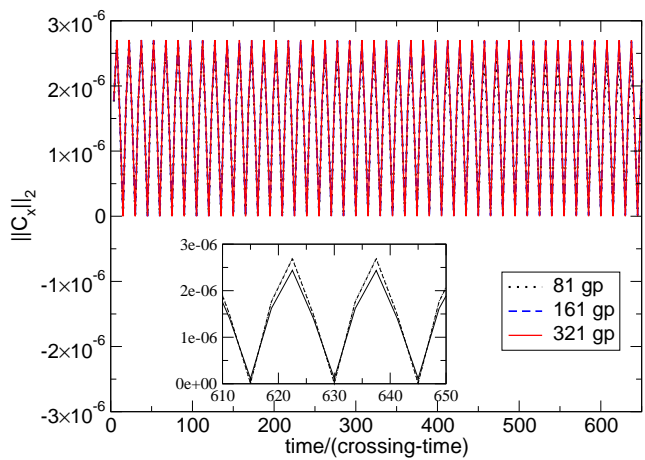} 
\caption{$L_2$ norm of $C_{x}$ vs. number of crossing times for three  
different resolutions, corresponding to the case where the initial 
data does not satisfy $C_x=0$. As expected from the continuum analysis, even in this case 
the norm does not grow in time. }
\label{gaugewavediss_CON_straight}  
\end{center} 
\end{figure}

The ALTI for $C_{{\cal A}}$ are, as in the previous examples, found  
to be of high frequency type and, therefore, the addition
of dissipation  
eliminates these growing modes. 
Figures~\ref{gaugewavediss_g_straight}--\ref{gaugewavediss_CON_straight}  
illustrate the behavior of simulations with the same initial data and boundary
conditions as those of Figures
\ref{gaugewave_C_straight}-\ref{gaugewave_g_straight} , 
except that now some dissipation is added, $\sigma=0.01$. The norm of 
$C_{\cal A}$ is illustrated in Fig.~\ref{gaugewavediss_C_straight}, where  now
only a slow growth is observed at late times.  Fig.
~\ref{gaugewavediss_CON_straight} shows, as before,  
the convergence factors (to zero) for $C_{\cal A}$, in the $L_2$ norm. 
Finally, figure \ref{gaugewavediss_g_straight} shows the $L_2$ norm of $g_{xx}$ for 
different resolutions. Only a linear growth is observed in the 
solution is found; furthermore, this growth decreases with resolution.

\begin{figure}[ht] 
\begin{center} 
\includegraphics*[height=6cm]{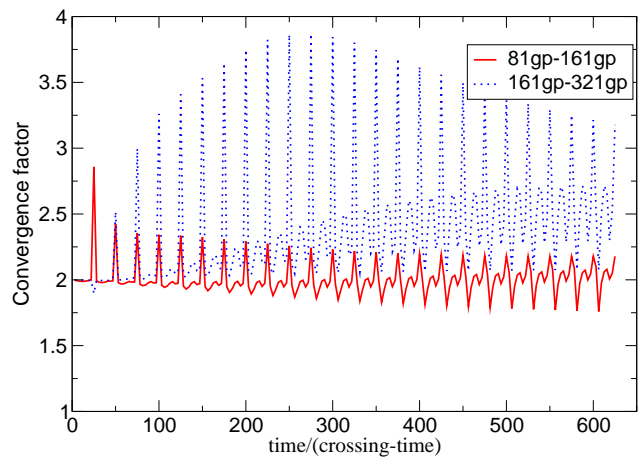} 
\caption{Convergence factors calculated with $C_{{\cal A}}$ vs. number of  
crossing times comparing runs with $81$ and $161$ points (dotted line), 
$161$ and $321$ points (solid line).}  
\end{center} 
\end{figure}

As a last example we consider initial data violating the constraints. Here,  
similarly good behavior is observed when dissipation is employed, in the sense
that, as predicted by the continuum analysis of Section \ref{eqs_gauge}, the constraints
oscillate in time. For instance,  
figure \ref{gaugewavediss_CON_straight} shows 
the $L_2$ norm of the constraint $C_x$ vs. time. As expected, these converge 
to a roughly constant behavior as resolution is increased, and no growth 
is observed.

%%%%%%%%%%%%%%%%%%%%%%%%%%%%%%%%%%%%%%%%%%%%%%%%%%%%%% 
\section{Conclusions}\label{conclusions} 
%%%%%%%%%%%%%%%%%%%%%%%%%%%%%%%%%%%%%%%%%%%%%%%%%%%%%% 
 
The construction of stable numerical methods for IBVPs often involves a 
mixture of intuition and experimentation.  The choice of numerical scheme 
and boundary conditions are often considered as necessarily separate ingredients.  
For certain problems, however, the 
energy method for discrete systems is an integrated, systematic method 
for creating numerically stable schemes and analyzing 
their behavior.  While this method has not yet been widely used in general 
relativity, many current outstanding problems in the field vitally depend 
on understanding the subtle interplay of boundaries with the interior 
computational domain.  In this paper we have applied the discrete energy 
method to test problems commonly found in numerical relativity and analyzed the numerical and 
long-term stability  of the resulting  
schemes, with and without black hole excision.  We have also   
investigated the numerical origin of fast growing modes which may 
artificially appear in  numerical solutions, and used the energy 
method to discuss possible remedies. 
 
In particular, we have presented explicit examples of the 
techniques described in paper I and discussed what leads to ALTI and 
a way to remove or alleviate them.  A message from this work is that 
it is not sufficient to construct numerically  stable implementations, but also 
special care must be taken to devise suitable {\em numerical} methods to rule out, or 
alleviate ALTI. (For a somewhat related approach, see the recent work \cite{winicourgauge}).  
 
To illustrate how this is achieved we have adopted a series of examples,  
which although 
considerably simpler than Einstein equations,  are relevant in the context of 
numerical relativity.  In particular,  in efforts to deal with black hole 
spacetimes and to understand the source of problems encountered with 
the gauge wave test.  As shown throughout this work, even though there are
examples in which low-frequency ALTI are present and
for which no sharp estimate is known, a careful analysis 
and grouping of the variables and/or the addition of 
(controllable) artificial dissipation may in many cases be crucial in handling otherwise 
artificial long term instabilities.

It is expected that these techniques 
will be important in a number of 
fronts,  in particular in the simulation of Einstein's equations. 
 
\section{Acknowledgments} 
We wish to thank G.~Calabrese, O.~Sarbach, J.~Pullin, E.~Tadmor, 
P.~Olsson, and H.-O.~Kreiss for helpful discussions, and E. Schnetter for
helpful comments on the manuscript.  This work was 
supported in part by NSF grants PHY0244699, PHY0312049; NASA-NAG5-1430; the Horace  
Hearne Jr.{} Lab for 
Theoretical Physics, CONICET, and SECYT-UNC. 
Computations were done at LSU's Center for Computation and 
Technology and parallelized with the CACTUS 
 toolkit~\cite{cactus}. L.L.{} is an Alfred P.{} Sloan Fellow.

\appendix 
%%%%%%%%%%%%%%%%%%%%%%%%%%%%%%%%%%%%%%%%%%%%%%%%%%%%%%%%%%%%%%%%%%%%%%% 
\section{\bf Cubic excision for Kerr black holes}  
\label{smallness} 
 
Black hole excision removes the singularity by placing an inner boundary 
on the computational domain.  While the boundary removes the singularity, 
the question now becomes finding physically and mathematically consistent inner boundary 
conditions.  
Clearly, a general solution to this problem for an arbitrary 
boundary is unknown.  The time-development of this boundary presents 
a further complication. Namely, if the boundary becomes time-like, it 
will encounter the singularity in a finite amount of time.  These two 
issues can be resolved simultaneously by adopting an inner boundary with 
a space-like world tube. This ensures it will not hit the singularity and 
that no boundary conditions are necessary if all the characteristics speeds
are within the light cone, as the past domain of dependence 
of boundary points is wholly contained on the previous time slice. Some 
freedom remains in choosing the particular region to be excised, and 
clearly the largest possible excision boundary is preferable to avoid 
the strongest gradients of the spacetime. 
 
To illustrate some of these issues we now examine cubical excision for both 
the Schwarzschild and Kerr black holes, looking for the {\it largest possible} 
space-like excision cube. 
As we discuss below, the {\it shape} of the boundary plays a crucial 
role in satisfying the spacelike condition. 
We examine  one face of the cube, a $x=\mbox{const}$ face. The condition 
that this surface is space-like implies that $g^{xx} < 0$, as $\nabla_x$ 
is the normal to the surface.  Since the coordinates are fixed {\it a-priori,} 
we can only vary the position of the boundary to meet this condition. 
We find that this condition is quite difficult to satisfy, and 
possible for only very slowly spinning black holes. 
 
To see this in detail, consider the Kerr metric in Kerr--Schild  
form~\cite{kerrschild}. The quantity $g^{xx}$ is 
\begin{equation} 
g^{xx} = 1 - 2 \frac{M r^3}{r^4+a^2z^2}  
         \left ( \frac{rx+ay}{r^2+a^2} \right )^2, 
\end{equation} 
where $r$ is related to the Cartesian coordinates by 
\begin{equation} 
\frac{x^2+y^2}{r^2+a^2}+\frac{z^2}{r^2} = 1 \, . 
\end{equation} 
The space-like character requirement on the surface implies that 
\begin{equation} 
2 \frac{M r^3}{r^4+a^2z^2} \left ( \frac{rx+ay}{r^2+a^2} \right )^2 > 1 \, . 
\end{equation} 
For the Schwarzschild black hole, $a=0$,  
$r^2=x^2+y^2+z^2$, and the above condition becomes 
\begin{equation} 
2 \frac{M x^2}{r^3} > 1.  
\end{equation} 
The corner of the cube is the limiting point ($y=z=x$) for the largest cube, 
giving the bound $x < 2M/(3\sqrt{3})$, or $x \lesssim 0.385 M$. 
 
\begin{figure}[ht] 
\begin{center} 
\includegraphics*[height=8cm,width=8cm]{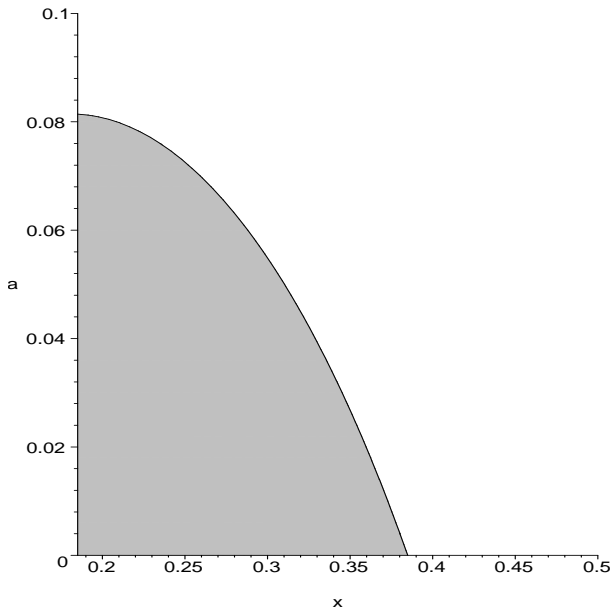} 
\caption{The shaded region indicates, for a given value of $a$,  
the size of the maximum allowed cubical region 
that satisfies Eq.~\ref{cubecond}. 
Clearly, the allowed values of $a$ are restricted below $\lesssim 0.082$.}. 
\label{kerr_excise}  
\end{center} 
\end{figure}

For the spinning black hole, the analysis is slightly more involved. The 
limiting point for the largest cube is given by $y=-x$, $z=\pm x$,  
giving the condition 
\begin{equation} 
\frac{ 2 M \sqrt{2} x^3 (\sqrt{3} x - {2}^{-1/2} a )^2}{ (3 x^2+a^2)^2 } > \frac{4 x^2 + a^2 2^{-1/2} }{2}.\label{cubecond} 
\end{equation} 
Examining this condition shows that cubical excision can not be used when 
$a\gtrsim 0.082$, as shown in Fig.~\ref{kerr_excise}. 
Furthermore, the singularity is not at a point, 
but is ring-like, implying that the cube may not be arbitrarily small. 
Finding the minimum sized cube is more involved, as the faces 
of the cube need to be examined in addition to the corners~\cite{gioeldave}. 
 
This limitation of cubical excision for the Schwarzschild spacetime was 
noted before~\cite{markITP}, by requiring that the excision 
boundary has no in-coming modes in a particular formulation of Einstein equations.  A similar 
analyzis to that of~\cite{markITP} for the Kerr spacetime and cubical excision regions 
has also recently  presented in~\cite{gioeldave}. Here we would like to point out  
the following. First, when  
the eigenmodes are complete and the speeds are physical (they lie within the null cone) 
those results and the one presented here are equivalent, as it is rooted in the 
physical nature of the excision boundary. Second, when dealing with a formulation 
admitting eigenspeeds outside the light cone, or a non-complete set of eigenmodes,   
the current straightforward analysis provides necessary conditions that
have to be met. These necessary conditions are useful 
when dealing with any formulation, in particular one  
whose eigenspeeds might not be known. 
Finally, we emphasize that these results are coordinate dependent; nevertheless, for the families 
often used in numerical relativity tests, the same result applies. In particular for  
the family of slicings reported by Martel and 
Poisson~\cite{martelpoisson} (which contains both the Kerr--Schild and 
Painlev\' e--Gullstrand slicings of Schwarzschild).

%%%%%%%%%%%%%%%%%%%%%%%%%%%%%%%%%%%%%%%%%%%%%%%%%%%%%% 

\end{document}